\documentclass[10pt,letterpaper]{article}

\usepackage[utf8]{inputenc}
\usepackage{cite}
\usepackage{booktabs}
\usepackage{natbib}
\usepackage{enumitem}
\usepackage{listings, color}
\usepackage{graphicx}
\usepackage{nameref,hyperref}
\newcommand{\eg}{e.\,g.,\ }
\newcommand{\ie}{i.\,e.,\ }

\definecolor{dkgreen}{rgb}{0,0.6,0}
\definecolor{gray}{rgb}{0.5,0.5,0.5}
\usepackage[ruled]{algorithm2e}

\SetAlFnt{\small}
\SetAlCapFnt{\small}
\SetAlCapNameFnt{\small}
\SetAlCapHSkip{0pt}
\IncMargin{-\parindent}

\begin{document}



\begin{flushleft}
{\Large
\textbf{Applying Cooperative Machine Learning to Speed Up the Annotation of Social Signals in Large Multi-modal Corpora}
}
\newline
\\
Johannes Wagner \textsuperscript{1,*},
Tobias Baur\textsuperscript{1},
Yue Zhang\textsuperscript{2},
Michel F. Valstar\textsuperscript{3},
Bj\"orn Schuller\textsuperscript{2},
Elisabeth Andr\'e\textsuperscript{1}
\\
\bigskip
\textsuperscript{1} University of Augsburg
\\
\textsuperscript{2} Imperial College London
\\
\textsuperscript{3} University of Nottingham
\\
\bigskip
* johannes.wagner@informatik.uni-augsburg.de
\end{flushleft}

\begin{abstract}

Scientific disciplines, such as Behavioural Psychology, Anthropology and recently Social Signal Processing are concerned with the systematic exploration of human behaviour. A typical work-flow includes the manual annotation (also called coding) of social signals in multi-modal corpora of considerable size. For the involved annotators this defines an exhausting and time-consuming task. In the article at hand we present a novel method and also provide the tools to speed up the coding procedure. To this end, we suggest and evaluate the use of Cooperative Machine Learning (CML) techniques to reduce manual labelling efforts by combining the power of computational capabilities and human intelligence. The proposed CML strategy starts with a small number of labelled instances and concentrates on predicting local parts first. Afterwards, a session-independent classification model is created to finish the remaining parts of the database. Confidence values are computed to guide the manual inspection and correction of the predictions. To bring the proposed approach into application we introduce \textsc{NOVA} -- an open-source tool for collaborative and machine-aided annotations. In particular, it gives labellers immediate access to CML strategies and directly provides visual feedback on the results. Our experiments show that the proposed method has the potential to significantly reduce human labelling efforts.

\end{abstract}

\section{Motivation}\label{sec:intro}

In various research disciplines (Behavioural Psychology, Medicine, Anthropology, ...) the annotation of social behaviours is a common task. This process includes manually identifying relevant behaviour patterns in audio-visual material and assigning descriptive labels. Generally speaking, segments in the signals are mapped onto a set of discrete classes, \eg a certain type of gesture, a social situation (\eg conflict), or the emotional state of a person. In Social Signal Processing (SSP) \cite{Pentland2007}, a subset of these events -- the so called \emph{social signals} -- are used to 
augment the spoken part of a message with non-verbal information to enable a more natural human-computer interaction \cite{Vinciarelli2008a,Vinciarelli2009}\footnote{To give an example of a social signal, think of a situation where we say something in a sarcastic voice to 
indicate that we actually mean the opposite.}. To automatically detect social signals from raw sensory input (\eg speech signals) machine learning (ML) can be applied. That is, sensory input is transformed into a compact set of relevant features and a classifier is trained on manually labelled examples to optimise a learning function. Once trained, the classifier can be used to automatically predict labels on unseen data.

However, since humans transmit non-verbal messages through a number of channels (voice, face, gestures, etc.\,) and due to the complex interplay between these channels (think, for instance, of a faked versus a real smile, which depends on subtle contractions of the muscles at the corner of the eyes as well as the timing of the muscle activations \cite{Valstar2007}), progress in SSP is directly linked to the availability of large and well 
transcribed multi-modal databases rich of human behaviour under varying context and different environmental settings \cite{Douglas-Cowie2003}. Common challenges in creating such datasets lie in the high degree of naturalness 
required of the recording scenarios, how well one recording scenario generalises to other settings, the number of human raters needed to reach a consensus on labels, and of course the sheer amount of data. Thinking of the many hours of labelled data that are required, it is clear that gathering large amounts of annotated training samples seems like an infeasible task, in respect to time, cost and effort.

Unfortunately, the 
bulk of databases that have been collected hitherto contain either acted behaviour recorded 
from a limited number of professional actors (\eg DaFEx \cite{Battocchi2005} or Emo-DB \cite{Burkhardt2005}) or isolated snapshots (\eg Belfast Naturalistic Database \cite{Douglas-Cowie2000} or VAM \cite{Grimm2008}). The proper training of an online recogniser, however, requires long and continuous recordings collected under preferably natural conditions \cite{Douglas-Cowie2007}. An example is the \textsc{SEMAINE} corpus \cite{McKeown2010}, which is composed of 100 sessions (each about 5 minutes) of emotionally coloured, yet free conversations. However, like other comparable corpora (\eg SAL \cite{Douglas-Cowie2008} or IEMOCAP \cite{Busso2008a}) it features audio-visual content only. Given the richness of observable social expressions there is still a growing need for labelled data of human interactions \cite{Pantic2007,Vinciarelli2012,Eerekoviae2014}. Not least because state-of-the art algorithms, such as deep neural networks (DNNs) profit from large amounts of labelled training data \cite{Sun:2017}.

Even though there exists a vast resource of raw data, which is nowadays pervasive in digital format and relatively easy and inexpensive to collect, \eg from public resources such as social media, the problem of gathering relevant annotations still needs to be overcome.
One approach is the \emph{Active Learning} (AL) \cite{Zhu2005} algorithm that interactively query the user to manually label certain data points. 
The core idea of AL is to extract the most informative instances from a pool of unlabelled data based on a specific query strategy \cite{Settles2009-ALL}. These selected instances are then passed to human 
annotators and finally -- after labelling -- a model is derived from this subset. This, of course, reduces the labelling effort. In addition, it has two more positive side effects. First, it speeds up the training since fewer instances have to be processed. Second, it helps improving the maximum accuracy, as it reaches a more coherent learning model (focussing on the most relevant cases). The work by \citet{Zhang2015a} takes the idea of AL a step forward and combines it with \emph{Semi-Supervised Learning} (SSL) techniques to efficiently share the labelling work between human and machine: a pre-existing classifier is used to derive confidence values for unlabelled data, thus human annotators are involved only for instances predicted with insufficient confidence. Such a strategy allows the performance of an existing classifier to be improved while minimising the costly work of human labelling. To further save labelling efforts one can apply \emph{Dynamic Active Learning} (DAL) by choosing the most reliable raters first \cite{Zhang2015c}. 
To put it into other words, the DAL algorithm also considers how many and which annotators should be queried on a per instance level.

\begin{figure}[t]
\centering
\includegraphics[width=1.0\textwidth]{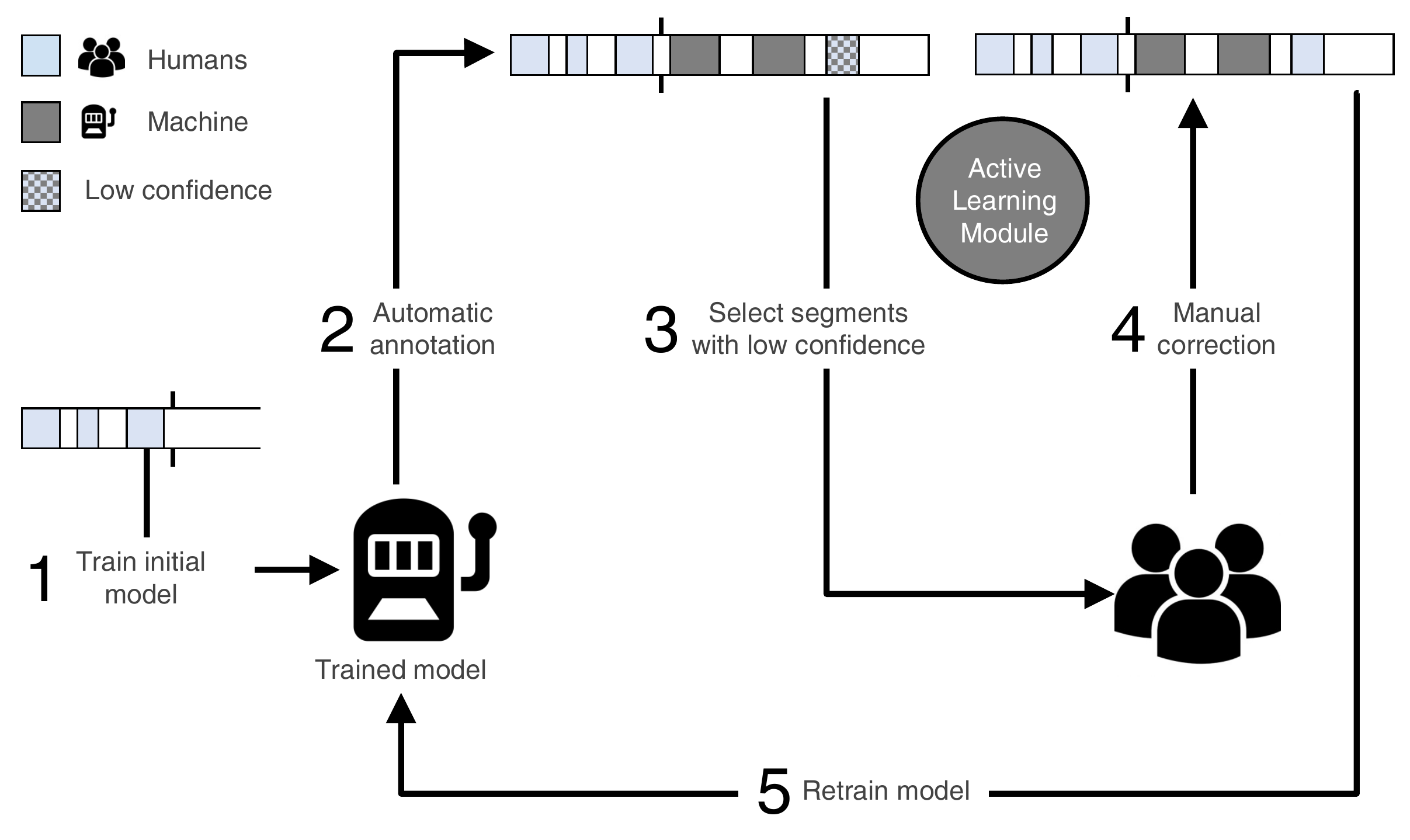}
\caption{The scheme depicts the general idea behind Cooperative Machine Learning (CML): (1) An initial model is trained on partially labelled data. (2) The initial model is used to automatically predict unseen data. (3) Labels with a low confidence are selected and (4) manually revised. (5) The initial model is retrained with the predicted / revised data.
}
\label{fig:cl:scheme}
\end{figure}

In this article, we subsume learning approaches that efficiently combine human intelligence with the machine's ability of rapid computation under the term \emph{Cooperative Machine Learning} (CML) \cite{Dong2003,Zhang2015a}. In Figure \ref{fig:cl:scheme} we illustrate our approach, which creates a loop between a machine learned model and human annotators: an initial model is trained (1) and used to predict unseen data (2). An active learning module then decides which parts of the prediction are subject to manual revision by human annotators (3+4). Afterwards the initial model is retrained using the new labelled data (5). Now the procedure is repeated until all data is annotated. By actively incorporating human expert knowledge into the learning process it becomes possible to interactively guide and improve the automatic prediction. Hence, the approach bears the potential to considerably cut down manual efforts. For instance, the system may quickly learn to label some simple behaviours, which already facilitates the work load for human annotators at an early stage. Then over time, it could learn to cope with more complex social signals as well, until at some point it is able to finish the task in a completely automatic manner. Such an iterative approach may even help bridging the gap between quantitative and qualitative coding, which still defines a great challenge in many fields in social science \cite{Chen2016}.

In the paper at hand, we aim at examining to what extent the proposed CML approach helps speeding up the annotation of social signals. In addition, we present a tool that allows researchers to apply the described techniques to their own databases. We see the main contributions of this work as follows:

\begin{itemize}

\item In Section \ref{sec:cl}, we propose a novel two-step CML strategy: as long as only few labelled instances are available the system is applied to local fractions of the database. Later, as more labelled instances become available, larger parts can be predicted.

\item In Section \ref{sec:evaluation}, we evaluate the proposed strategy on an audio-based annotation task by simulating the incremental injection of additional information during training. Results show that the proposed strategy significantly reduces manual coding efforts.

\item In Section \ref{sec:nova}, we introduce an open-source tool for collaborative and machine-aided labelling (\textsc{NOVA}). A walk-through is presented to demonstrate the collaborative annotating capability of the system. Experiences from users working with the tool are reported and discussed in Section \ref{sec:experiences}.

\end{itemize}

For the sake of clarity related work will be given separately for each section.



\section{Cooperative Machine Learning}\label{sec:cl}

Interactive machine learning \cite{Fails:Olsen:2003,Amershi:et:al:2014} aims to involve users actively in the creation of models for recognition tasks. Most approaches integrate automated data analysis and interactive visualisation tools in order to enable users to inspect data, process features and tune models. In this section, we focus on approaches that facilitate the acquisition of annotated data sets and introduce a novel methodology for applying \emph{Cooperative Machine Learning} (CML) to speed up annotation of social signals in large multi-modal databases.

\subsection{Related Work}

A common approach to reduce human labelling effort is the selection of instances for manual annotation based on active learning techniques. The basic idea is to forward only instances with low prediction certainty or high expected error reduction to human annotators \cite{Settles:2012}.
An art of its own right thereby is how to estimate which are these most informative ones. A whole range of options to choose from exist, such as calculation of `meaningful' confidence measures, detecting novelty (\eg by training auto-encoders and seeing for the deviation of input and output when new data runs through the auto-encoder), estimating the degree of model change the data instance would cause (\ie seeing whether knowing the label of a data point would make a change to the model at all), or trying to track `scarce' instances, \ie trying to find those data instances that are rare in terms of the expected label.
Further more sophisticated approaches aggregate the results of machine learning and crowdsourcing processes to increase the efficiency of the labelling process. \citet{Kamar2012} made use of learned probabilistic models to fuse results from computational agents and human labellers. They showed how to allocate tasks to coders in order to optimise crowdsourcing processes based on expected utility. \citet{Zhang2015d} developed an agreement-based annotation technique that dynamically determines how many human annotators are required to label a selected instance. The technique considers individual rater reliability and inter-rater agreement to decide on a combination of raters to be allocated to an instance. Active learning has shown great potential in a large variety of areas including document mining \cite{Tong:Koller:2002}, multimedia retrieval \cite{Wang:Hua:2011}, activity recognition \cite{Stikic:et:al:2008} and emotion recognition \cite{Zhang2015b}.

Most studies in this area focus on the gain obtained by the application of specific active learning techniques. However, little emphasis is given to the question of how to assist users in the application of these techniques for the creation of their own corpora. While the benefits of integrating active learning with annotation tasks has been demonstrated in a variety of experiments, annotation tools that provide users with access to active learning techniques are rare. Recent developments for audio, image and video annotation that make use of active learning include CAMOMILE \cite{Poignant2016} and iHEARu-PLAY \cite{Hantke2015}. However, systematic studies focusing on the potential benefits of the active learning approach within the annotation environment from a user's point of view have been performed only rarely \cite{Cheng:2015,Kim:2017}.

While techniques that enable systems to learn from human raters have become widespread, little attention has been paid to usability challenges of the remaining tasks left to end-users \cite{Amershi:et:al:2014}. \citet{Rosenthal2010} investigated which kind of information should be provided to users in order to reduce annotation errors in a setting for active learning. They found out that contextual information and predictions of the learning algorithms were in particular useful for the annotation of activity data. In contrast, uncertainty information had no effect on the accuracy of the labels, but just indicated to the labellers that classification was hard. \citet{Amershi2009} investigated how to empower users to select samples for training by appropriate visualisation techniques. They found that a representative overview of best and worst matching examples is of higher value than a set of high-certainty images and conjecture that high-certainty images do not provide much information to the learning processing due to their similarity to already labelled images. In another paper by \citet{Amershi2015} the authors suggest an interactive visualization technique to assess model performance by sorting samples according to their prediction scores. In their tool the user can directly inspect samples to retrieve additional information and annotate them for better performance tracking. This way, the tool allows users to monitor the performance of individual samples while the model is iteratively retrained.

The approaches above supported users in the annotation and selection of samples for training. As an alternative, graphical user interfaces have been developed that enable users to create their own annotated examples for training models. Typically, the labels are given by instructions or stimuli to be provided to the users to evoke particular behaviours. An example includes SSI/ModelUI \cite{Wagner2010}. It presents users with a graphical user interface that allows them to test different machine learning algorithms on labelled data. Labels are acquired by stimuli which may include textual instructions, but also images or videos. However, users have to determine themselves which kind of stimuli and data are most useful to create and tune models.

Summing up, it may be said that many studies experientially investigate the potential of novel techniques to minimise human labour.
In addition, few studies were run to actually label novel data, rather than test whether such method could save effort. Also note that the prevailing choice is merely active learning rather than the combination with semi-supervised learning, \ie cooperative machine learning.
Relatively little attention has been paid, however, to the question of how to make these techniques available to human labellers. There is a high demand for annotation tools that integrate cooperative machine learning in order to reduce human effort –- in particular in the area of social signal processing where human raters typically disagree on the labels \cite{Lotfian2017}.
In such a setting, dynamic active or cooperative strategies appear particularly promising, \ie not only learning the target task, but also as much as possible about the raters and their reliability depending on the labels and the content being labelled. Likewise, it can be learned `whom to trust when' to further reduce annotation effort by only requiring labels from the `right persons at the right time'.

\subsection{Two-fold Strategy}
\label{sec:twofold}

The cooperative machine learning strategy we propose here is a two-fold one. It is divided into a \emph{Session Completion} (SC) step during which information of a fraction of a single session is used to complete the remaining part of the session, and a \emph{Session Transfer} (ST) step during which information from a set of labelled sessions is used to predict a set of unlabelled sessions. In our understanding, a session defines a single continuous and self-contained recording. The sessions of a database can be captured on different dates and sites involving different subjects.

The division is motivated by the lack of labelled data in the beginning of an annotation process, which usually does not allow to build models that are robust enough to generalise well to the unseen parts. This is especially true if the recording conditions and the involved subjects vary between the individual sessions. Nevertheless, already small fractions of labelled data can be sufficient to build models that are able to make reliable predictions on data that resembles the instances that have been seen so far. An example is data recorded from the same subject under comparable conditions -- something we can generally expect from snapshots of the same session. Even if these models are too ``weak'' to make reliable predictions for the whole dataset, they can help to speed up the early annotation process. In the following, we refer to a classifier trained on samples of a single session as a \emph{session-dependent} classifier. Once enough sessions have been completed, a \emph{session-independent} model can be trained and used to accomplish remaining sessions.

To ensure the quality of the recognition, manual verification of the outcome of the classification might be necessary. This procedure can be accelerated by rating the predictions, \eg by adding confidence values to the predicted instances. Instead of reviewing everything annotators can concentrate on parts with \emph{low confidence}, \ie labels that have been predicted with a high uncertainty\footnote{In a multi-class classification task uncertainty can \eg be derived from the distance a predicted sample has to the decision boundaries of the other classes.}. The proposed strategy can be summarised as follows:

\begin{enumerate}

\item \textbf{Session Completion}: Manually assign labels to a fraction of a session and train a session-dependent classifier. Apply it to complete the remaining fraction. Based on the confidence values generated by the model query manual revision.

\item \textbf{Session Transfer}: Take all (with aid of step 1) fully labelled sessions and train a session-independent classifier. Apply it to predict annotations for remaining sessions. Again, based on the generated confidence values decide which parts require manual adjustment. 

\end{enumerate}

So far we have distinguished between session-dependent and session-independent classification. Depending on the corpus to which the strategy is applied, this may not necessarily be the best practise. For instance, if a dataset is composed of recordings that are too short to apply the first step we can adapt the strategy and initially complete recordings belonging to the same subject. Once we have labelled data from a sufficient number of individual subjects, we continue by training a subject-independent model and apply it to the remaining recordings. Likewise, we can use the described strategy across several databases, too. In that case we would concentrate on individual databases first and afterwards obtain a database-independent model that we use to label the remaining databases.

\subsection{Implementation}\label{sec:implementation}

To efficiently apply the described strategy, we would like to know the \emph{sweet spot} for applying the \textit{Session Completion} and the \textit{Session Transfer} step. On the one hand, if we apply it too early the model becomes unstable and predictions will be poor. On the other hand, if we annotate more data than necessary we give away precious time. To avoid any of the described situations, we are interested in finding a good trade-off between machine performance and human effort. Unfortunately, we cannot easily guess what is the ideal moment to hand over the task to a machine. This is because the amount of training data that is required to build a robust model depends on a number of factors, such as the homogeneity of the data, the discrimination ability of the extracted features, the number of subjects and classes, and not least at the complexity of the recognition problem. Alternatively, instead of trying to determine a sweet spot beforehand (and possibly miss it), we could iteratively test the applicability of the strategy and stop when the performance seems promising.

Therefore, we opted to make the described strategy an integral part of our tool (see Section \ref{sec:nova}). This allows annotators to visually examine the results at any time and to individually decide whether more labelling is required or not. However, this means that the time it takes to run the CML strategy becomes a crucial factor. Generally, it should not take longer than a few seconds or the annotation process will be interrupted (this is especially true for the session completion step). To reach this goal, we should reuse as much information as possible. One possibility is to apply classification on a small sliding window (frames) and use a rather simple (\eg linear) classifier. The former means that features have to be extracted only once (or can be even pre-extracted); the latter ensures a fast training.

\begin{figure}[t]
\centering
\includegraphics[width=0.75\textwidth]{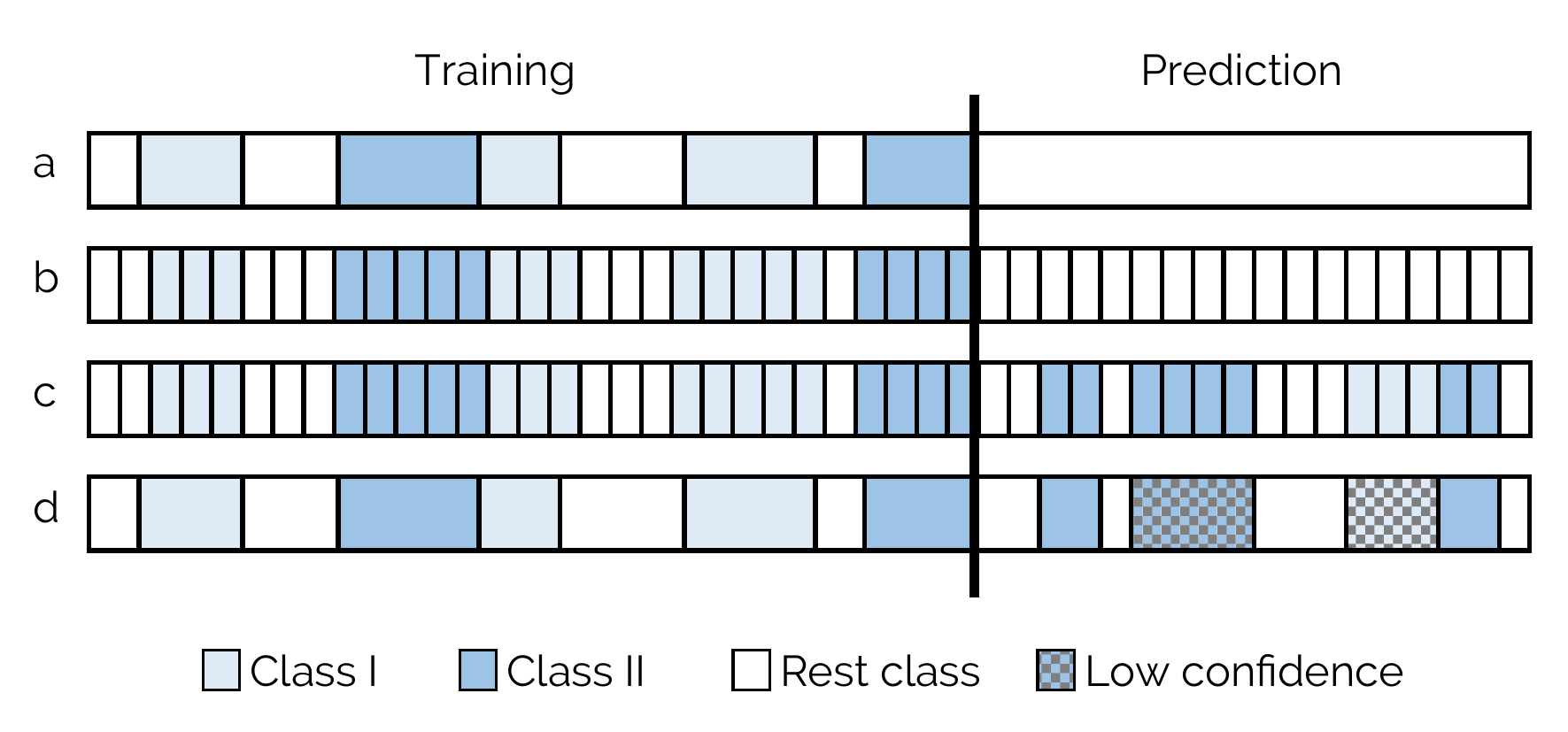}
\caption{Visualisation of the cooperative machine learning strategy by means of the SC step: (a) the end point of the last segment of the manual annotation defines where the training fraction ends and prediction begins. (b) labelled segments are mapped onto frames and empty frames are assigned to a rest class. (c) a model is build from the frames in the training fraction and used to predict the frames in the prediction fraction. (d) successive frames with the same class label are combined, the rest class is removed and segments with a low confidence are highlighted.}
\label{fig:cl:training}
\end{figure}

In the following, we restrict ourselves to complete discrete annotations, \ie we deal with multi-class problems. In case of the SC step we receive the raw signal stream (\eg an audio signal) of the current session and a partly finished annotation composed of labelled segments with a discrete start and end point. The segments can be of variable length and there may be gaps between two successive segments. By applying the following procedure we then predict the segments for unlabelled fraction of the session (see also Figure \ref{fig:cl:training}):

\begin{enumerate}
\item If not provided, extract frame-wise features for the whole session.
\item Find the frame that coincides with the end point of the last label in the annotation and split the feature sequence into a training fraction (preceding frames) and a prediction fraction (successive frames).
\item In the training set assign frames that overlap with a labelled segment by at least 50\% to the corresponding class. In case of several candidates keep the dominant one (most overlap). Assign remaining frames to a rest class.
\item Learn a classifier using all frames from the training fraction.
\item Use the classifier to label the prediction fraction by assigning to each frame the class with the highest probability.
\item Combine successive frames belonging to the same class and keep the average probability of the combined frames as confidence. Remove frames that belong to the rest class. Optionally, apply thresholds to remove small segments and fill gaps.
\item Add the predicted segments to the original annotation and mark segments with a low confidence.
\end{enumerate}

The ST step works in the same way with the difference that whole sessions are used to train the classifier, which is applied to predict whole sessions afterwards.


\section{Evaluation}\label{sec:evaluation}

We now turn to some experiments in which we examine the practical effect of the proposed Cooperative Machine Learning (CML) strategies of Section \ref{sec:cl}. We do this by means of a database including natural human-human interaction and simulate a situation where the detection system is applied to predict unlabelled fractions of the dataset. Using the original and predicted parts of the corpus to train a final detection model we evaluate the robustness and efficiency of the CML approach.

\subsection{Database and Problem Description}\label{sec:database}

\begin{figure}[t]
\centering
\includegraphics[width=1.0\textwidth]{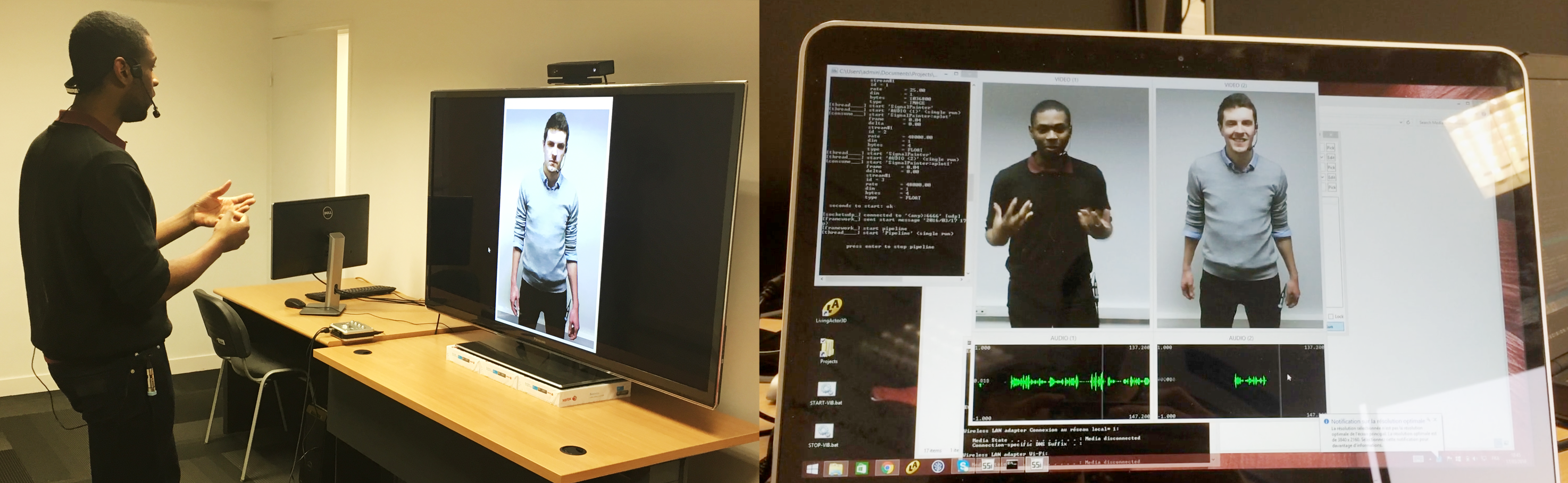}
\caption{Snapshots of user interaction (left) and observer screen (right) during a recording session for the NOXI database.}
\label{fig:noxi:setup-5}
\end{figure}

Let us quickly introduce the database and the classification problem we choose as a testbed for applying the proposed CML strategies. The NOvice eXpert Interaction database (\textsc{NOXI}) database \cite{Cafaro:2017} is a corpus of screen-mediated face-to-face interactions that features natural interactions between human dyads in an expert-novice knowledge sharing context. In a session one participant assumes the role of an expert and the other participant the role of a novice. The corpus was created as a part of the ARIA-Valuspa \cite{valstar2016} (Artificial Retrieval of Information Assistants – Virtual Agents with Linguistic Understanding, Social skills, and Personalised Aspects) project, and therefore has a strong focus on medial face-to-face interactions in an Expert-Novice setting. Figure \ref{fig:noxi:setup-5} shows two users during interaction.

One purpose of \textsc{NOXI} is to study interruption strategies. For instance, when a listener decides to ask a question or comment to what the speaker was saying and therefore starts an attempt to take over the speech turn. The simplest way to detect such situations is by looking for spots where the voice of the two participants is overlapping. If afterwards a speaker change occurs we can assume that the interrupting party successfully took over the turn. Otherwise we can treat it as a failed attempt. However, an interposed utterance is not necessarily a signal to interrupt the speaker. It can also be an expression of approval or interest, denoted as \emph{backchannels}. Likewise, not every speaker pause signals a floor change if, for instance, the speaker needs time to think what to say next. To bridge these pauses speakers usually utter a \emph{filler} sound. Hence, to correctly identify speaker interruptions we have to separate backchannels and fillers from other speech parts.

In the following, we present a detection system that is trained to automatically identify backchannels and fillers in speech. First, we evaluate the system following a classic machine learning approach to measure the performance of the system. Afterwards, we examine if and to what extent the system is able to speed-up the manual annotation process in the CML loop.

\subsection{Related Work}


Spoken dialogs do not exclusively contain the information that is being exchanged. A significant part of the spoken content is dedicated to the management of the communication and the expression of nuances of attitude and intention \cite{Yngve1970}. Generally, we distinguish between \emph{backchannels} and \emph{fillers}. In English, backchannels are utterances such as ``uh-huh'' and ``hmm'', as well as, short words such as ``yeah'', ``okay'', ``really?', etc. Their purpose is to verbalise active listenership and assessments, as well as requests for clarification or sentence completions \cite{Young2004}. Generally, they actively support the speaker to keep the floor. Likewise, fillers are sounds or words that speakers use during a pause to signal that they are not yet finished talking. When a listener hears a filler he continues listening rather than start talking. For instance, speakers interpose fillers when they search for a word or take a pause to think about what to say next. Common filler sounds in the English language are ``um'', ``er'' or ``ah'' \cite{Clark2002}. Like backchannels they help to continue conversations smoothly. In addition to backchannels and fillers, laughter defines a third group of vocalisations that are often interposed by both, the speaker and the listener, without the intention to invoke a turn change. Laughter is often a sign of amusement or joy, though it may also
show scorn or embarrassment \cite{Ruch2001}.

Previous work has shown that backchannels and fillers can be detected from the acoustic and the phonetic properties of speech. \citet{Edlund2010} detect backchannels using a K-Nearest Neighbors classifier on duration and inter-speaker relative loudness. Their system distinguishes backchannels from other vocabulary with an overall accuracy of 73\%. \cite{Prasad2010} disambiguate the use of the Hindi word ``h\~a'' -- which is both a discourse particle and also as a lexical equivalent of ``yes'' -- into its three discourse functions: questions, backchannels and agreement. They extract features based on pitch contour, power and duration, to which they apply a k-Means clustering. In their system, backchannels are correctly identified with an accuracy of 65\%. The detection of fillers and laughter events has been picked up by the INTERSPEECH 2013 Computational Paralinguistics Challenge \cite{Schuller2013}. The acoustic feature set proposed by the organizers of the challenge is derived from energy, cepstral and voicing related low-level descriptors and their deltas. The final feature vector combines the arithmetic mean and standard deviation across the current frame (at a frame step of 10 ms) and eight of its neighbouring frames. Recognition accuracy is reported as the Area Under the Curve measure \cite{Witten2005} yielding an unweighted average (UAAUC) of 82.9\% for the laughter classes and 83.6\% for the filler classes after training a Support Vector Machines (SVM) classifier. By adding phonetic features results can be improved by another $5\%$ \cite{Wagner2013a}.

A number of studies have investigated the use of Hidden Markov Models (HMMs) for modelling the temporal variations of different laughter sounds. For example, \citet{Tanaka2011} investigate four phonetic categories of laughter (nasal, voiced, chuckle, or ingressive) based on their spectral features. With this approach, the single laughter categories are recognised with an accuracy rate of 86.79\%. \citet{Schuller2008} test various classifiers using spectral features to distinguish four kinds of isolated non-verbal vocalisation (laughter, breathing, hesitation, and consent). In their experiment, HMMs achieve an accuracy rate of 92.3\% and outperform Hidden Conditional Random Fields (HCRFs) and Support Vector Machines (SVMs). \citet{Scherer2012} show that the performance of a laughter classifiers depends on whether it is applied in online or offline mode. SVMs gave the best results for offline mode with given segmentations. However, in online mode, they were surpassed by Echo State Networks (ESNs) and HMMS because they fail to find on- and offsets of laughter. A particular challenge in laughter detection is overlapping laughter in conversational speech. \citet{Truong2012} conducted a study using four corpora in order to identify spectral features that differentiate overlapping and non-overlapping laughter.

\subsection{Detection System}\label{sec:detection}

Though in our experiments we concentrate on the detection of speech and fillers/ backchannels, we opt for a detection system that is as generic as possible. This will allow us to apply it to other classification problems, too. Also, speed performance plays a crucial role as we do not want to interfere with the annotation process. In the following we start by describing the proposed generic detection system.

Due to its modularity and capability of fast online incremental processing we rely on the \textsc{OpenSmile} audio feature extraction tool \cite{Eyben2013}. However, we refrain from using a large statistical feature set like the ComParE (Computational Paralinguistics Evaluation) set, which assembles 6\,373 features by brute-force combination of Low Level Descriptors (LLDs) with Functionals \cite{Schuller2013}. This kind of feature sets are usually applied on chunks of several seconds length (\eg a whole utterance). In our scenario, however, we opt for a frame-based feature set extracted over a small moving window that can be reused across successive training steps. Also, we should keep in mind that especially in the beginning of the annotation process the size of the training sets can be small. In that case a smaller feature set will lower the risk of overfitting.

\begin{figure}[t]
\centering
\includegraphics[width=0.75\textwidth]{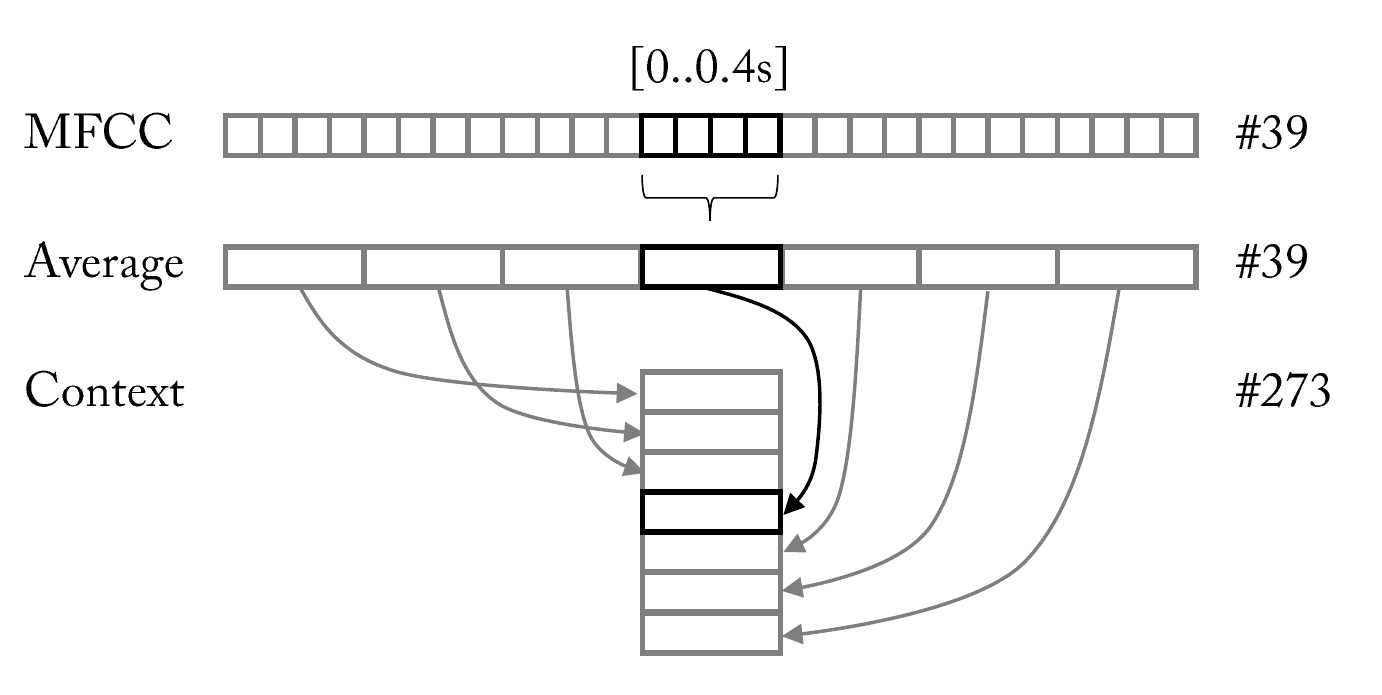}
\caption{Illustration of the feature extraction step. First, four MFCC frames with a dimension of 39 are averaged to reduce the sample rate of the signal to 25 Hz. Afterwards, neighbouring frames are added, here 3 frames from the left and 3 frames from the right. This results in a final feature vector of size 273.}
\label{fig:vad:feature}
\end{figure}

Mel-Frequency Cepstral Coeffcients (MFCC) provide a compact representation of the short-term power spectrum. Not only have they a long tradition in speech recognition systems \cite{Rabiner1993} and speaker verification tasks \cite{Ganchev2005}, but have also been successfully applied in the field of social signal processing, \eg emotional speech recognition \cite{Lee2004,Vogt2005,Beritelli2006,Neiberg2006,Schuller2007,Kishore2013} and laughter detection \cite{Kennedy2004,Knox2007,Urbain2010}. For our tests, we calculate 13 Mel-Frequency Cepstral Coeffcients (including the 0$^{th}$ coefficient) and their first- and second-order frame-to-frame difference (delta-delta). According to standard practice we use a moving window of 25 ms with a frame step of 10 ms. Afterwards we reduce the stream to a frame step of 40 ms by averaging always four frames. This ensures that the sample rate of the feature stream is consistent with the video frame rate of 25 Hz. Though not relevant for the current study, it will be handy if we want to integrate visual features in the future. Yet, 40 ms are small enough to detect start and end point of voiced segments sufficiently accurate. Since the length of the filler events we want to detect may be longer than 40 ms, we optionally concatenate neighbouring frames from both sides of the current frame -- in the following denoted as context size $n$. A context of size 3, \eg means that the current frame is extended by 3 frames from the left and 3 frames from the right. This increases the number of features by a factor of $2 \cdot n + 1$. Figure \ref{fig:vad:feature} illustrates the feature extraction step.



As classification model we use a linear Support Vector Machine (SVM) provided by \textsc{LIBLINEAR} -- a Library for Large Linear Classification \cite{Fan2008}. Since the implementation does not use kernels, training time is significantly reduced even for large input sets composed of several ten thousand samples. For multi-class classification we select a L2-regularised logistic regression solver (option -s 0) and add a bias term of 0.1 (option -B 0.1). We keep default values for all other parameters. Since we expect unbalanced class distributions, we randomly remove samples to match the size of the class with the least number of samples. Finally, features values are scaled between -1 and 1 (when we test a sample we apply the scaling derived from the training set). Confidence values are scaled in a way such that individual class scores sum to 1.

\subsection{Results}\label{sec:results}

Having established a generic classification system we will now evaluate recognition performance on the \textsc{NOXI} corpus (see Section \ref{sec:database}). We pick 18 sessions (German sub-corpus) and randomly split them into a training set including two-third of the sessions 
summing up to nearly 7 h of audio data. The remaining 6 sessions
form the test set with an overall duration of almost 3.5 h.

\begin{figure}[t]
\centering
\fbox{\includegraphics[width=0.75\textwidth]{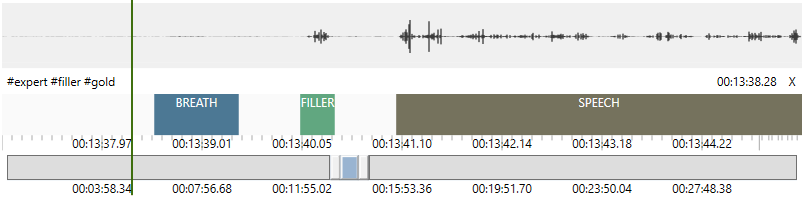}}
\caption{Example of a manual annotation.}
\label{fig:vad:annotation}
\end{figure}

To evaluate the proposed detection system we need to establish a ground truth. We use \textsc{NOVA} (will be introduced in Section \ref{sec:nova}) to manually annotate voiced parts in the audio files. To not introduce a machine bias none of the CML strategies described in Section \ref{sec:cl} are applied. Manual annotation is accomplished by three experienced annotators\footnote{Two research assistants who have been working in the field of SSP for several years and one master student who took part in an annotation course.}, each completing six sessions. Table \ref{tab:vad:statistic} lists the applied annotation scheme. Since labels are assigned to voiced sounds the remaining parts implicitly define the rest class \textsc{SILENCE}. Because of the better audio quality we use the head set recordings. However, it turned out that the close-talk recordings tended to pick up breathing sounds, so we introduce an additional \textsc{BREATH} class to prevent false alarms during silenced parts. Backchannels, fillers, laughter, and other voiced sounds such as grunts and coughs, are gathered in a single class denoted as \textsc{FILLER}. Speech segments that are neither backchannels nor fillers are labelled as \textsc{SPEECH}. An example of an annotation is shown in Figure \ref{fig:vad:annotation}. We asked the raters to measure how long it took to annotate the sessions. In total they spent a little more than 14 h, which results in an average time of 47 minutes per session.

Next, we split the annotations in frames of 40 ms length and extract MFCC features, which results in 946 783 frames (exact class distribution are given in Table \ref{tab:vad:statistic}). We sample the training set down to 22 918 samples per class and train a linear SVM model. Results are summarised in Table \ref{tab:vad:result-1}. We report classwise recognition accuracy and Unweighted Average (UA) recall (average across classes). For a direct comparison with the INTERSPEECH 2013 Social Signals Paralinguistic Challenge we also consider the Area Under the Curve (AUC) measure. A 85\% AUC for the \textsc{FILLER} class (best case) shows that results are comparable to \citet{Schuller2013}. We take this as a prove that our detection system does a reasonable job on the examined task.

\begin{table}[t]
\caption{Annotation scheme and frame number per class.\label{tab:vad:statistic}}
{
\begin{tabular}{llrlrl}
\toprule
Class & Description  & Train & \% & Test & \% \\
\midrule
SPEECH & Speech (except filler and backchannels) & 265 466 & 41.4 & 126 183 & 41.2 \\
BREATH & Breathing (except unvoiced laughter) & 22 918 & 3.6 & 3 929 & 1.2 \\
FILLER & Backchannels, fillers, laughter, & 26 665 & 4.2 & 8 592 & 2.8 \\
 	   & and other voiced sounds & & \\
\emph{SILENCE} & Implicit rest class representing unvoiced parts & 325 528 & 50.8 & 167 502 & 54.7 \\
\midrule
$\sum$ & & 640 577 & & 306 206 & \\
\bottomrule
\end{tabular}
}
\end{table}

\begin{table}[t]
\caption{Classwise recall and Area Under the Curve (in brackets) in \% with respect to the context $n$. UA = Unweighted Average, UAAUC = UA of AUC\label{tab:vad:result-1}}
{
\begin{tabular}{rcccccc}
\toprule
$n$          & 0 & $1$ & $2$ & $5$ & $10$ & $15$ \\
\midrule
SPEECH     & 64.7 (95) & 67.6 (96) & 69.5 (96) & 73.7 (97) & 74.6 (97) & 74.3 (97) \\
BREATH     & 82.5 (95) & 84.3 (96) & 85.1 (97) & 87.2 (98) & 87.9 (98) & 88.2 (98) \\
FILLER     & 46.6 (69) & 54.1 (74) & 59.1 (77) & 66.1 (82) & 71.9 (84) & 74.1 (85) \\
SILENCE    & 82.9 (92) & 83.1 (93) & 83.9 (94) & 85.5 (95) & 84.0 (96) & 82.8 (96) \\
\midrule
UA (UAAUC) & 69.2 (88) & 72.3 (90) & 74.4 (91) & 78.1 (93) & 79.6 (94) & 79.8 (94) \\
\bottomrule
\end{tabular}
}
\end{table}

\begin{figure}[t]
\centering
\includegraphics[width=0.75\textwidth]{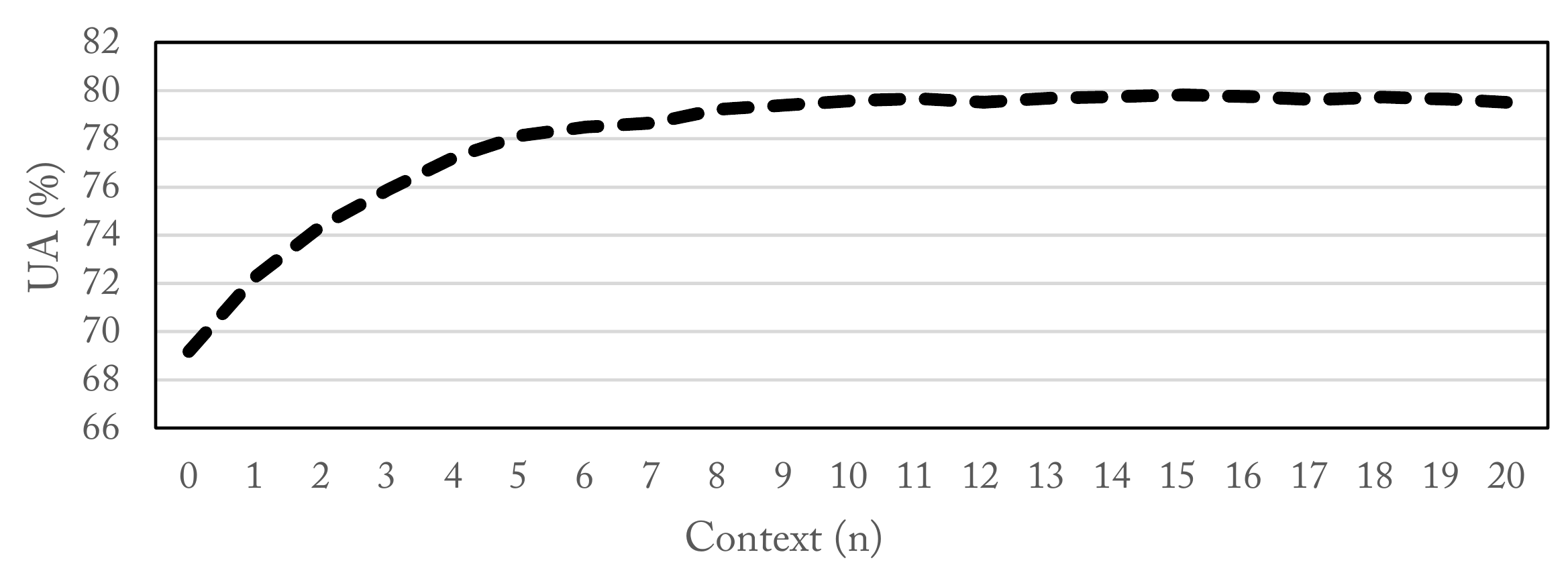}
\caption{Classwise UA recall in \% with respect to the context size $n$.}
\label{fig:vad:context}
\end{figure}

As seen in Table \ref{tab:vad:result-1} increasing the number of concatenated frames has a positive effect on the recognition accuracy ({\raise.17ex\hbox{$\scriptstyle\mathtt{\sim}$}}10\%). Especially the FILLER class benefits from a larger frame context (25\% improvement), which we explain with the fact that fillers are usually short and isolated speech episodes surrounded by silence. In Figure \ref{fig:vad:context} we notice a saturating effect for more than 10 frames. 

To give an impression how the system performs in terms of speed we report measurements on an Intel(R) Core(TM) i7-3930K. In our tests extracting MFCC-based features with a context of size 5 and a frame step of 0.04 s took 0.9 s for one minute of mono audio sampled at 48 kHz. Extrapolated to 10.5 h of interaction it requires less than 10 minutes to extract features for the whole German subset. Since features are reused this defines a one-time effort. Training a linear classifier on the training set (91 672 frames after class balancing) took on average 50 s. Frame-wise prediction on the test set (306 206 frames) only {\raise.17ex\hbox{$\scriptstyle\mathtt{\sim}$} 2.9 s. Such values suggest that the proposed detection system is fast enough to be embedded into the annotation process without causing serious interruptions (even if several hours of data are used as input / output).

\subsection{CML Simulation}\label{sec:simulation}

Finally, we want to know how the proposed detection system performs in combination with the proposed CML strategies. In Section \ref{sec:cl} we have defined the \emph{sweet spot} as the moment when additional annotation efforts no longer improve the stability of the classification model. Practically, this defines the ideal point to hand the task over to the machine. To experientially determine the sweet spot for the given problem, we incrementally inject information into the training process. In the following, we simulate this procedure by splitting the original training set into two parts: we assume that $n$ sessions have been manually labelled (subset $L$), whereas the remaining sessions are yet unlabelled (subset $U$). Now, we derive three classifiers $c$, $c'$ and $c''$ (see Figure \ref{fig:vad:evaluation}):

\begin{description}[labelindent=1pt,style=multiline,leftmargin=0.75cm]
\item [$c$] Train with the labels of $L$.
\item [$c'$] Use $c$ to predict the labels of $U$ and retrain with the predicted labels.
\item [$c''$] Before retraining inspect the predicted labels if their confidence is below a threshold $t$ and correct them if necessary.
\end{description}

\begin{figure}[t]
\centering
\includegraphics[width=0.75\textwidth]{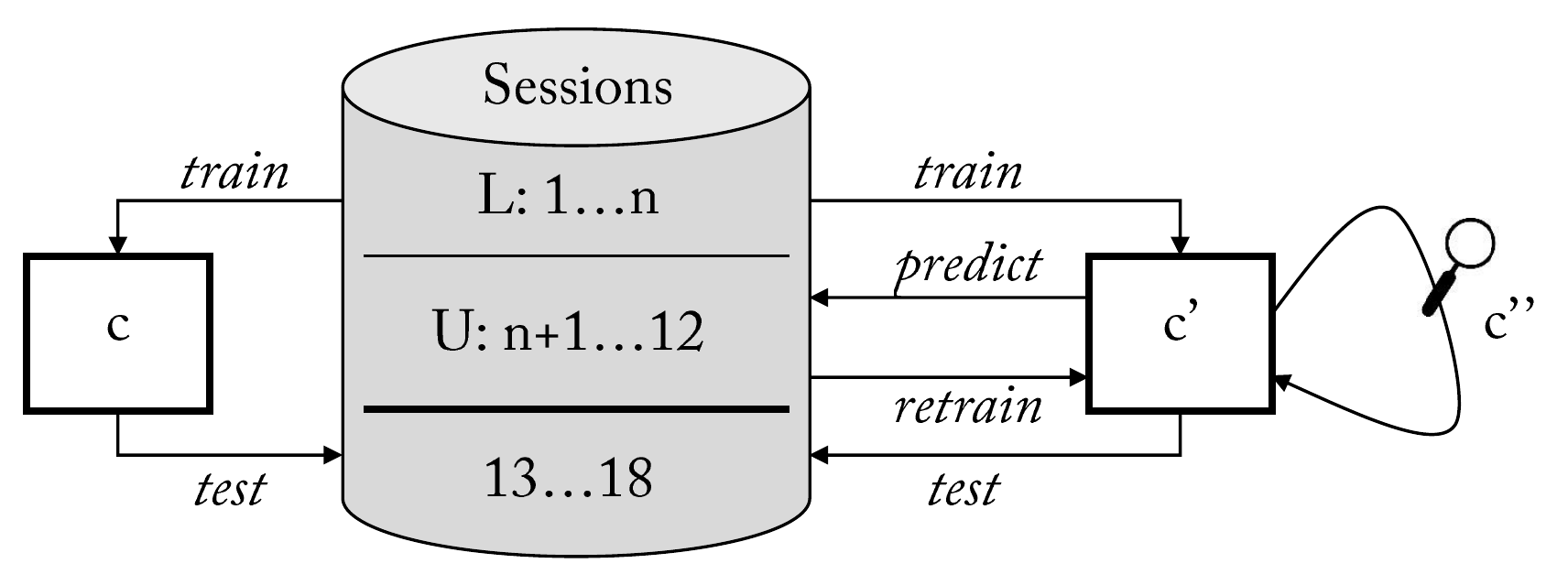}
\caption{In the default condition a classifier $c$ is evaluated after training with labelled sessions (L) only. In case of $c'$ unlabelled sessions (U) are predicted and used to retrain the model. And in case of $c''$ predicted labels are reviewed and possibly corrected before retraining takes place.}
\label{fig:vad:evaluation}
\end{figure}

$c'$ simulates the case where the annotation process is stopped at some point and the labelled fraction of the database is used to predict the remaining parts. Note that in this case \emph{all} predicted labels are included during the final training step, i.\,e.\ no automatic selection strategies and no additional manual efforts are applied.

$c''$ simulates the case where parts of the prediction are inspected (here the selection is based on the class confidence). To assess the additional manual effort we measure what we call the \emph{Inspection Rate} (IR), which is the fraction of frames below the confidence, and the \emph{Correction Rate} (CR), which is the fraction of frames that are finally assigned a different label.

\begin{table}[ht]
\caption{Recognition results on the test when incrementally injecting information into the training process using the three classifiers $c$, $c'$, $c''$ (see remarks in text). In case of $c''$ $t$ defines the confidence threshold for inspecting predicted labels. In each row we start with $n$ labelled sessions. Results are obtained with the detection system described earlier using a stacking context of 5.\label{tab:vad:result-2}}
{
\begin{tabular}{r|c|c|ccc|ccc}
\toprule
 &	$c$ & $c'$  & \multicolumn{3}{c|}{$c''$ ($t=0.5$)} & \multicolumn{3}{c}{$c''$ ($t=0.75$)} \\			
\midrule
$n$	&	UA (\%)	&	UA (\%)&	UA (\%)	&    IR      &    CR      & UA (\%)   &         IR &    CR \\
\midrule
1	&	67.2	&	70.1	&	74.1	&	38\%	&	14\%	&	\textbf{77.4}	&	\textbf{87\%}	&	\textbf{25\%}  \\
2	&	72.3	&	70.5	&	74.3	&	51\%	&	17\%	&	78.0	&	82\%	&	25\%  \\
3	&	73.0	&	71.6	&	76.0	&	36\%	&	12\%	&	77.9	&	66\%	&	18\%  \\
4	&	74.4	&	73.2	&	76.4	&	36\%	&	12\%	&	78.0	&	59\%	&	18\%  \\
5	&	76.2	&	75.8	&	76.9	&	31\%	&	11\%	&	78.0	&	51\%	&	16\%  \\
6	&	76.3	&	76.4	&	\textbf{77.1}	&	\textbf{27\%}	&	\textbf{9\%}     &	78.0	&	45\%	&	14\%  \\
7	&	76.4	&	76.4	&	77.2	&	25\%	&	9\%    	&	78.2	&	40\%	&	13\%  \\
8	&	77.0	&	76.3	&	78.0	&	15\%	&	5\% 	&	78.0	&	26\%	&	7\%   \\
9	&	76.8	&	76.9	&	78.1	&	11\%	&	4\%	    &	78.1	&	19\%	&	6 \%  \\
10	&	\textbf{77.8}	&	\textbf{77.8}	&	78.0	&	5\%	    &	2\%	    &	77.9	&	10\%	&	2\%   \\
11	&	78.1	&	77.9	&	78.1	&	1\%	    &	0\%	    &	78.1	&	4\%	    &	1\%   \\
12	&	78.1	&	78.1	&	78.1	&	-	    &	-	    &	78.1	&	-    	&	-   \\
\bottomrule
\end{tabular}
}
\end{table}

Table \ref{tab:vad:result-2} summarises the performance of $c$, $c'$ and $c''$ on the test set (the same as before). In each row we assume that $n$ sessions of the original training set have been labelled (\eg $n=4$ means that $L$ consists of sessions 1 to 4 and $U$ consists of sessions 5 to 12). Based on the results we can gain some interesting insights. Let us therefore assume we aim for a classification model that is at maximum one percent worse than the reference model trained on all sessions, i.\,e.\ has an Unweighted Average (UA) recall of at least 77.1\% (throughout the tests we have applied a stacking context of 5).

The performance of classifier $c$ shows that ten of the twelve sessions are sufficient to yield a 77.8\% recognition accuracy. Hence, to achieve our goal we can stop after labelling ten sessions and skip the last two. Now, what happens if we extend the training set with predicted labels (no selection or manually correction yet)? Checking the results of $c'$ we see that again ten sessions are required to achieve the desired accuracy. In fact, extending the training set with purely predicted data generally has no positive effect on the recognition performance. Although disappointing at first glance this is actually not too surprising. Obviously we cannot expect to improve a model unless we inject some new knowledge, which is not the case if we add predictions without inspection. This is as if we asked a student to revise his own test, which is pointless unless we point out some of his mistakes first.

Hence, some manual efforts are needed here. And indeed: after correcting frames with a confidence below 0.5 (that is 9\% of all frames in the remaining subset) $c''$ yields 77.1\% already after 6 sessions. To achieve this we actually had to review 27\% of predicted frames. If we assume that the remaining six sessions make up approximately half of the frames this corresponds to $\frac{1}{8}$ of the full training set, i.\,e.\ in total we have to examine $\frac{5}{8}$ (= $\frac{1}{2} + \frac{1}{8}$) of the training data. As mentioned earlier the average time to annotate a session was 47 minutes. Hence, we can reckon a saving of approximately 3.5 hours (5.9 h instead of 9.4 h). Obviously, this significantly speeds up the annotation process.

Apparently, the more work we are willing to spend on the correction of predicted labels the earlier we receive a stable classification model. In fact, if we lift the correction threshold to 0.75 we observe that $c''$ now yields 77.4\% already after the first session. However, this is achieved at the expense of a more than three times higher inspection rate (87\%), which means that we have to view almost $\frac{9}{10}$ of the corpus (precisely $0.87 \frac{9}{10} + \frac{1}{10}$). Hence, it can be a better strategy to complete a couple of sessions first and in return apply a smaller correction threshold afterwards leaving less data for inspection (more on that in Section \ref{sec:experiences}).


\section{NOVA Tool}\label{sec:nova}

The results of the previous section encouraged us to integrate the proposed Cooperative Machine Learning (CML) approach into our annotation tool \textsc{NOVA} ((Non)Verbal Annotator). This way we give annotators the possibility to immediately inspect and if necessary correct predicted annotations. Though an earlier version of the tool existed (see \cite{Baur2015}), we extended it to achieve a seamless integration of the collaborative annotation process. \textsc{NOVA} is open-source and can be downloaded from \url{http://github.com/hcmlab/nova}.

\subsection{Related Work}

\textsc{NOVA}'s interface has been inspired by existing annotation tools. For instance, EUDICO Linguistic Annotator (\textsc{ELAN}) \cite{Wittenburg2006}, Annotation of Video and Language (\textsc{ANVIL}) \cite{Kipp2013}, and \textsc{EXMARALDA} (Extensible Markup Language for Discourse Annotation) \cite{Schmidt2004}. These tools offer layer-based tiers to insert time-anchored labelled segments, that is \emph{discrete} annotations. \emph{Continuous} annotations, on the other hand allow an observer to track the content of an audiovisual stimulus over time based on a continuous scale. A tool that allows labellers to trace emotional content in real-time on two dimensions (activation and evaluation) is \textsc{FEELTRACE} \cite{Cowie2000}. Its descendant \textsc{GTRACE} (General Trace) \cite{Cowie2012} allows the user to define their own dimensions and scales. Other tools to accomplish continuous descriptions are \textsc{CARMA} (Continuous Affect Rating and Media Annotation) \cite{Girard2014} and \textsc{DARMA} (Dual Axis Rating and Media Annotation) \cite{Girard2016}. An interesting approach for gathering crowd-sourced annotations is \textsc{iHEARu-PLAY} \cite{Hantke2015}, that allows labelling audio material on a valence-arousal scale in form of a browser-game. Whereas most tools are restricted to describe audiovisual data by a single user, \textsc{repoVizz} \cite{Mayor2013} is an integrated online system to collaboratively annotate streams of heterogeneous data (audio, video, motion capture, physiological signals, etc\.,). Datasets are stored in an online database, allowing users to interact with the data remotely through a web browser.

Though the mentioned tools are of great help to create annotations at a high level of detail, the tools offer none or only little automation. Since labelling of several hours of interaction is an extremely time consuming task, methods to automate the coding process are highly desirable. To this end \textsc{NOVA} has been advanced with features to create collaborative annotations and to apply CML strategies out of the box (see Section \ref{sec:cl}). To support a truly collaborative work-flow between several annotators and the machine a database back-end is provided to store, exchange, and combine annotation work.

\subsection{General Interface}

\begin{figure}[t]
\centering
\fbox{\includegraphics[width=0.98\textwidth]{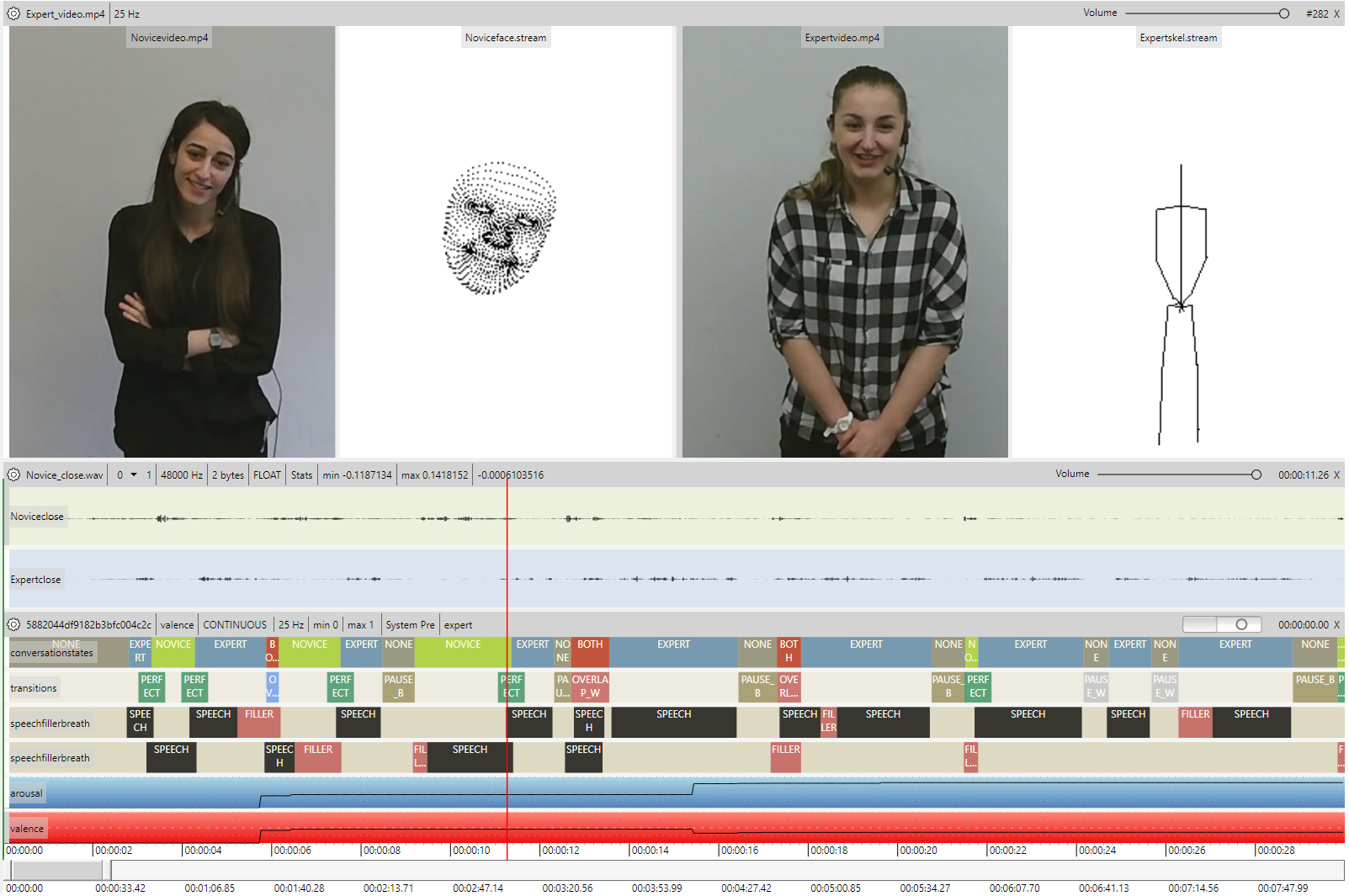}}
\caption{\textsc{NOVA} allows it to visualise various media and signal types and supports different annotation schemes. From top down: full-body videos along with skeleton and face tracking, and audio streams of two persons during an interaction. In the lower part several discrete and continuous annotation tiers are displayed. Annotations can be edited on a static fraction of the recording or interactively during playback.}
\label{fig:nova:overall}
\end{figure}

\begin{figure}[t]
\centering
\vspace{1cm}
\includegraphics[width=1.0\textwidth]{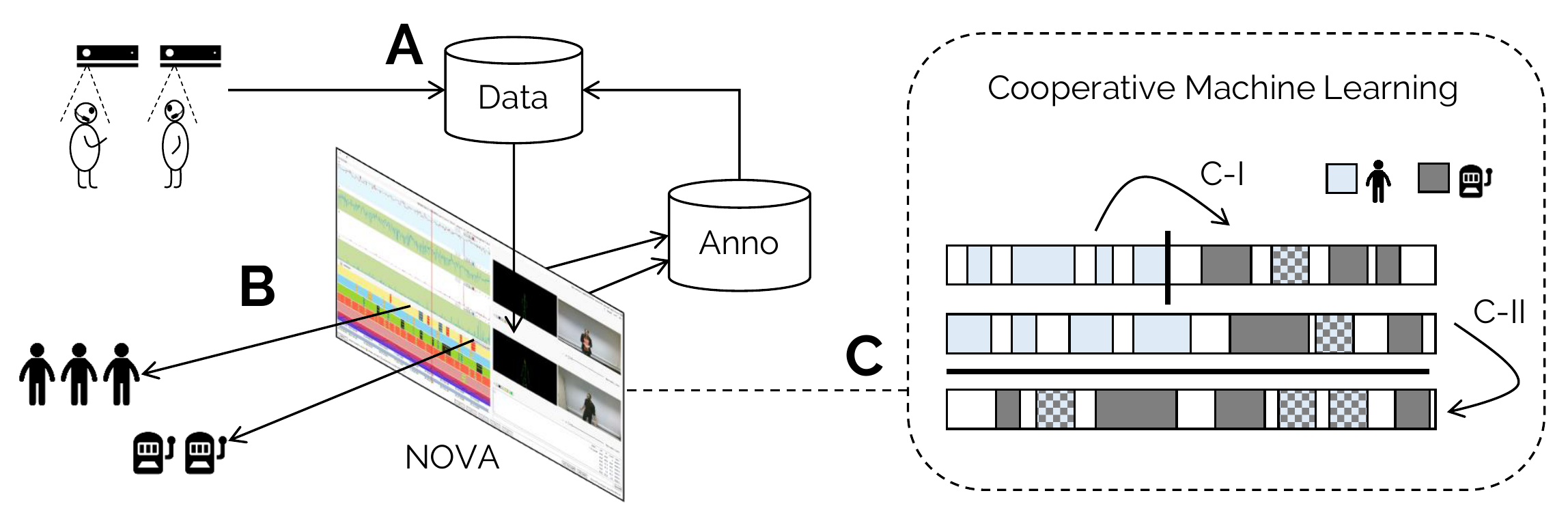}
\caption{CML integration in NOVA: (A) A database is populated with recordings of human interaction. (B) NOVA functions as interface to the data and provides a database to distribute and accomplish annotation tasks among human annotators. (C) At times, CML is applied to automatically complete unfinished fractions of the database: (C-I) A session-dependent model is trained on a partly annotated session and applied to complete it. (C-II) A pool of annotated sessions is used to train a session-independent model and predict labels for the remaining sessions. In both cases, confidence values guide the revision of predicted segments (here marked with a pattern).
}
\label{fig:cl:scheme-old}
\end{figure}

The NOVA user interface has been designed with a special focus on the annotation of long and continuous recordings involving multiple modalities and subjects. Unlike other annotation tools, the number of media files that can be displayed at the same is not limited and various types of signals (video, audio, facial features, skeleton, depth images, etc.) are supported. Further, multiple types of annotation schemes (discrete, continuous, transcriptions, geometric, etc.) can be selected to describe the visualised content (see Figure \ref{fig:nova:overall} for an example). Several statistics are available to process the annotations created by multiple coders. For instance, statistical measures such as Cronbach's $\alpha$ \cite{Cronbach:1951} or Cohen's $\kappa$ \cite{Cohen:1960} can be applied to identify inter-rater agreement and annotations from multiple raters can be merged from the interface.

In the following we will concentrate on another feature of \textsc{NOVA}: the use of CML tools to speed up the annotation of multi-modal corpora. Figure \ref{fig:cl:scheme-old} shows \textsc{NOVA} as the mediator between a database and several human and machine annotators. Both CML steps described in Section \ref{sec:cl} are supported by the interface.



\subsection{Database Back-end}

To support a collaborative annotation process, \textsc{NOVA} maintains a database back-end, which allows users to load and save annotations from and to a MongoDB\footnote{\url{https://www.mongodb.com/}} running on a central server. This gives involved annotators the possibility to immediately commit changes and follow the annotation progress of the others. MongoDB is an open-source and cross-platform NoSQL database. We have chosen it in favour of a relational database due to its simplicity and fast read/write operations.

\begin{figure}[t]
\centering
\includegraphics[width=1.0\textwidth]{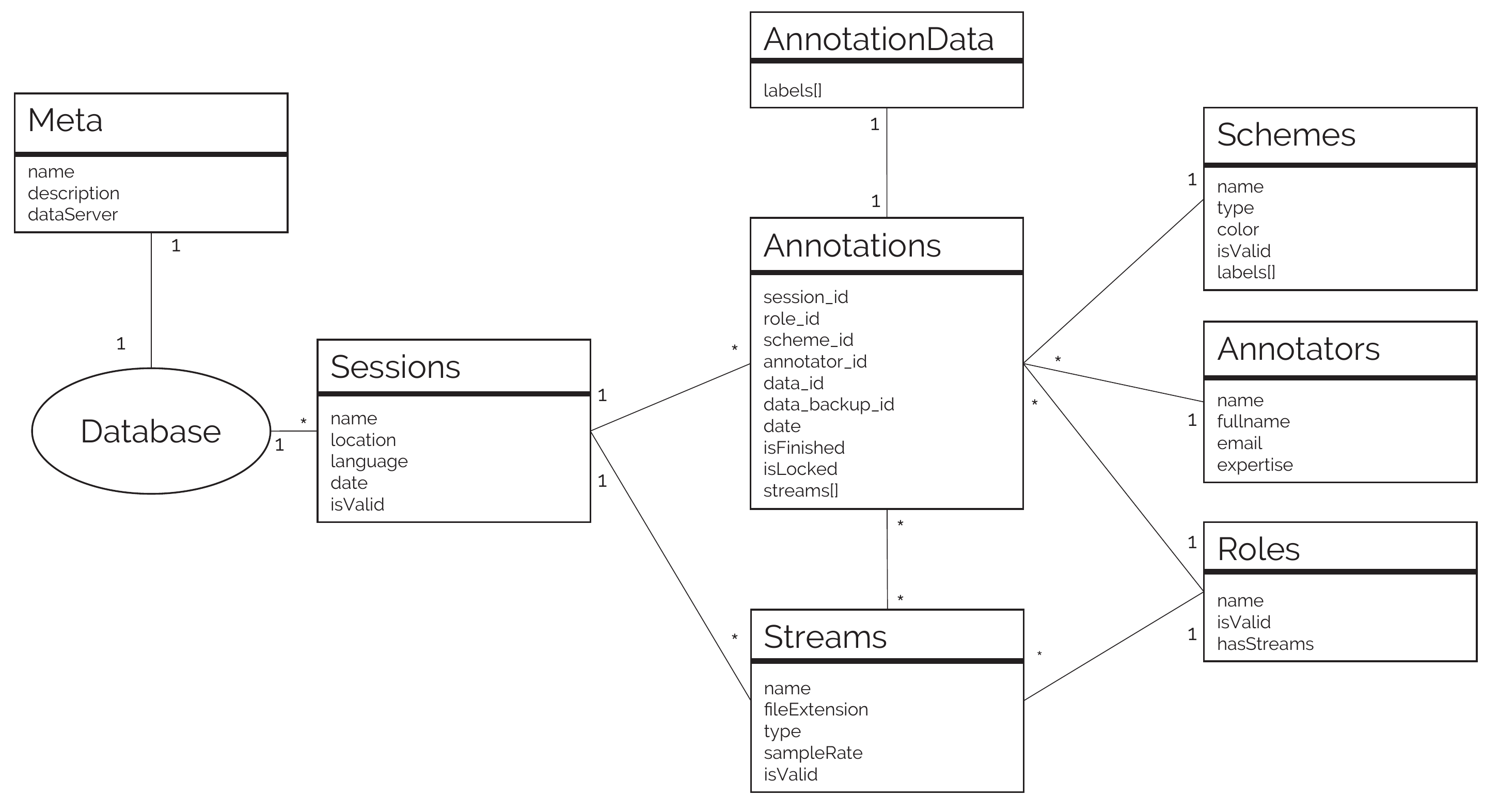}
\caption{Overview of NOVA's database structure. Annotations and meta information on subjects, sessions, etc. are stored in different collections. NOVA includes necessary tools to maintain and populate a database.}
\label{fig:nova:database}
\end{figure}

We opt for a design that not only allows to read and write annotations, but manages all relevant meta data of a corpus, too. Generally, each corpus is represented by a single database including several collections (the analogous to tables in relational databases). The collections are (see also Figure \ref{fig:nova:database}):
\begin{itemize}
\item \textbf{Meta}: Meta information about a database, including the data server location, and a description
\item \textbf{Sessions}: Stores general information for each recording session, such as location, language and date.
\item \textbf{Annotators}: Stores names and meta information of the involved annotators (human or machine!).
\item \textbf{Roles}: Stores the different roles subjects can take on during a recording session (\eg listener vs speaker).
\item \textbf{Streams}: Stores the recorded stream files. Each file is assigned to a media type, a session, a subject and a role. An url is included that points to the location where the file can be downloaded.
\item \textbf{Schemes}: Stores the available annotation schemes.
\item \textbf{Annotations}: Stores the headers of created annotations. An annotation is linked to an annotator, an annotation scheme, a role and a session. Optionally, a list of stream files is referenced to store which information should be displayed during the annotation process.
\item \textbf{AnnotationData}: Contains the actual annotation data (segments or scores) for an annotation. Additionally a Backup is stored for each annotation, allowing the user to go back to the previous version.
\end{itemize}

As soon as several users collaborate on a common database it becomes crucial to implement adequate security policies. For instance, we want to prevent a situation in which a user accidentally overwrites the annotation of another user. Therefore, standard users can only edit and delete their own annotations. They can, however, load annotations of other users. In that case the annotation is copied and stored under their username. Only users, privileged with admin rights may edit and delete annotations of other users. They can also assign newly created annotations to specific users. This way, an admin can divide up forthcoming annotation tasks among the pool of annotators.

Beside human annotators, a database may also be visited by one or more ``machine users''. Just like a human operator they can create and access annotations. Hence, the database also functions as a mediator between human and machine. To control the annotation progress we have introduced a `isFinished' flag that signals if an annotation requires further fitting or is finished. A second flag `isLocked' marks whether an annotation is editable or not.

\textsc{NOVA} provides instruments to create and populate a database on a MongoDB server from scratch. This gives users the possibility to apply the tool on their own corpora. At any time new annotators, schemes and additional sessions can be added. No specific knowledge about databases is required.

\subsection{ML Back-end}

For best possible performance tasks related to machine learning (ML) are outsourced and executed in a background process. As ML framework we use our open-source Social Signal Interpretation (\textsc{SSI}) framework\footnote{\url{http://openssi.net}}. \textsc{SSI} has been successfully applied to a couple of recognition problems in the past, see \eg \cite{Urbain2010,Wagner2011,Lingenfelser2011,Wagner2013c,Lingenfelser2014,Wagner2015b}. Since \textsc{SSI} is primarily designed to build online recognition systems, a trained model can be directly used to detect social cues in real-time \cite{Wagner2013b}.

Though, \textsc{SSI} is developed in C++, it offers a simple XML interface to define feature extractors and classifiers. For instance, the definition of the MFCC features from Section \ref{sec:detection} looks as follows:

\begin{lstlisting}
<chain>
  <meta frameStep="10ms" rightContext="15ms"/>
  <filter>
    <item create="OSPreemphasis"/>
  </filter>
  <feature>
    <item create="OSMfccChain" option="mfccdd"/>
  </feature>
</chain>
\end{lstlisting}
\vspace{-0.5cm}

When applied to a stream, the signal values are first run through a pre-emphasis filter before MFCC features are extracted over a sliding window of 25 ms with a frame step of 10 ms (timings can be overwritten in \textsc{NOVA}). To configure the MFCC extraction (\eg the number of coefficients) a separate option file is created (here 'mfccdd'). However, \textsc{SSI} supports other features sets, too. For instance, it allows to run scripts from the widely used \textsc{OpenSmile} toolkit \cite{Eyben2013}. And it provides feature sets for other type of signals. For instance, a wrapper for the \textsc{OpenFace} tool \cite{Baltruvsaitis:2016} is available to extract of facial points and action units from video streams.

Likewise, the classification model from Section \ref{sec:detection} is defined as follows:

\begin{lstlisting}
<trainer>
  <meta balance="under"/>	
  <normalize>
    <item method="Scale"/>
  </normalize>
  <model create="LibLinear" option="linsvm"/>
</trainer>
\end{lstlisting}
\vspace{-0.5cm}

Here, \textsc{SSI} is configured to balance the number of class samples by removing samples from overrepresented classes and scale features into a common interval. As training model a linear SVM will be used. However, \textsc{SSI} also supports other classification models such as Google's neural network framework \textsc{TensorFlow}\footnote{https://www.tensorflow.org/} or the popular \textsc{Theano}\footnote{https://github.com/Theano/} library.

\subsection{CML Walk-through}\label{sec:nova:walkthrough}

We will conclude this section with a walk-through that demonstrates NOVA's CML tools. We assume that a database has been created and populated with several sessions of audio recordings from one or more users. In our case, we work on the \textsc{NOXI} database described in \ref{sec:database} and apply the annotation scheme used during the evaluation in Section \ref{sec:results}, \ie we want to mark filler and breath events in regular speech by assigning the labels \textsc{BREATH}, \textsc{FILLER} and \textsc{SPEECH}. Note that number and names of the classes is defined by the underlying annotation scheme, which be easily adapted by the user to fit any other labelling problem.

\begin{figure}[t]
\centering
\fbox{\includegraphics[width=\textwidth]{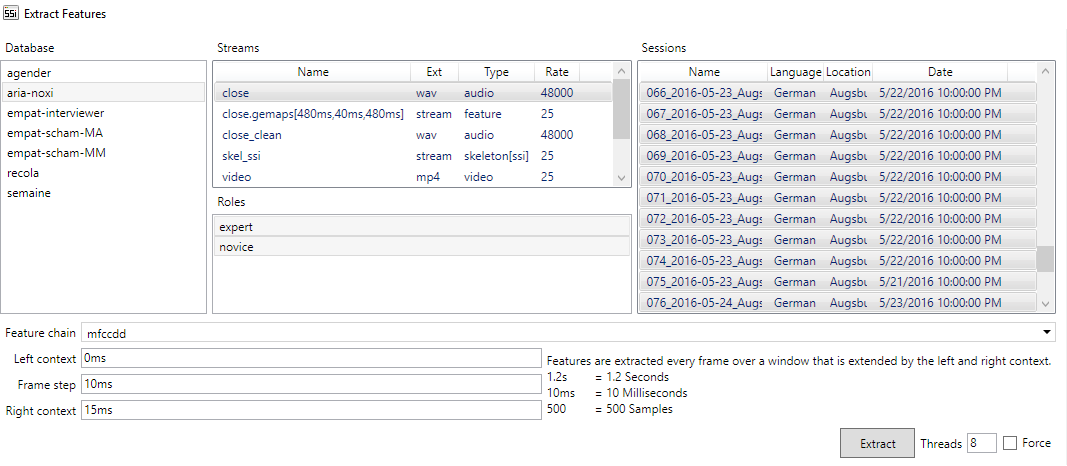}}
\caption{Screenshot of the feature extraction dialogue. The user chooses a stream (here audio) and an according feature extraction method (here mfccdd). Feature extraction is applied for the selected roles and sessions.}
\label{fig:nova:feature}
\end{figure}

As a first step, we extract MFCC features for the German sessions in the \textsc{NOXI} database. The dialogue is shown in Figure \ref{fig:nova:feature}. It allows us to choose a source stream and a feature extraction method (only methods that can be applied to the selected stream will be listed). Optionally, we can overwrite the default frame step and context sizes. Extraction can be accelerated by running several sessions in parallel (here 8).

\begin{figure}[t]
\centering
\fbox{\includegraphics[width=1.0\textwidth]{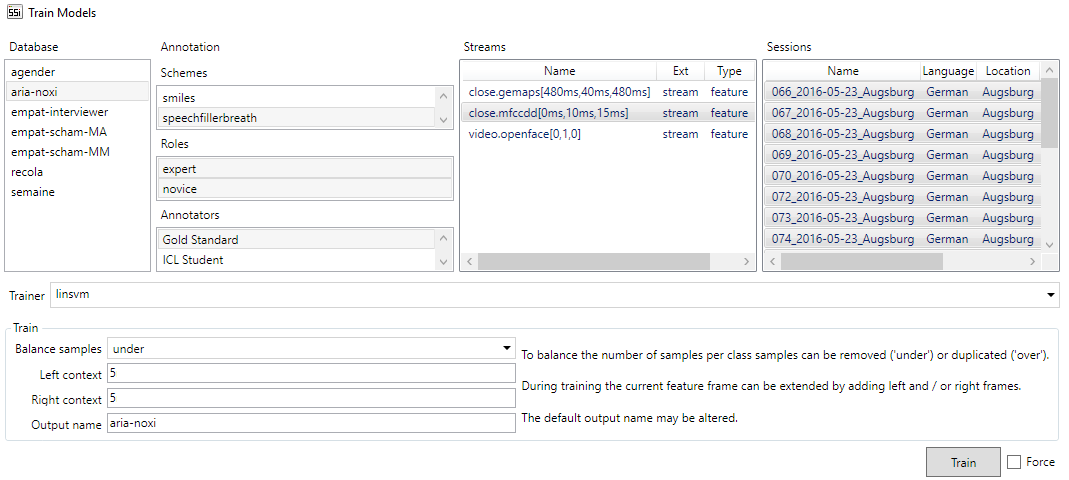}}
\caption{Screenshot of the model training dialogue. The user selects a coding scheme, a role and an annotator (here Gold Standard). Sessions for which an according annotation exists are now displayed and a stream can be selected to define the input for the learning step. Finally, a model (here linsvm) is chosen and the training begins.}
\label{fig:nova:training}
\end{figure}

In a next step, we can now pick an annotation scheme and apply it to the previously extracted feature streams. Figure \ref{fig:nova:training} shows the interface that allows us to select the input and choose a classification model (only models are shown that fit the selected input). Optionally, we can set a left and right context to concatenate neighbouring feature frames (see Section \ref{sec:detection}). Afterwards the trained model is stored and can now be applied to predict unlabelled data.

\begin{figure}[ht]
\centering
\fbox{\includegraphics[width=\textwidth]{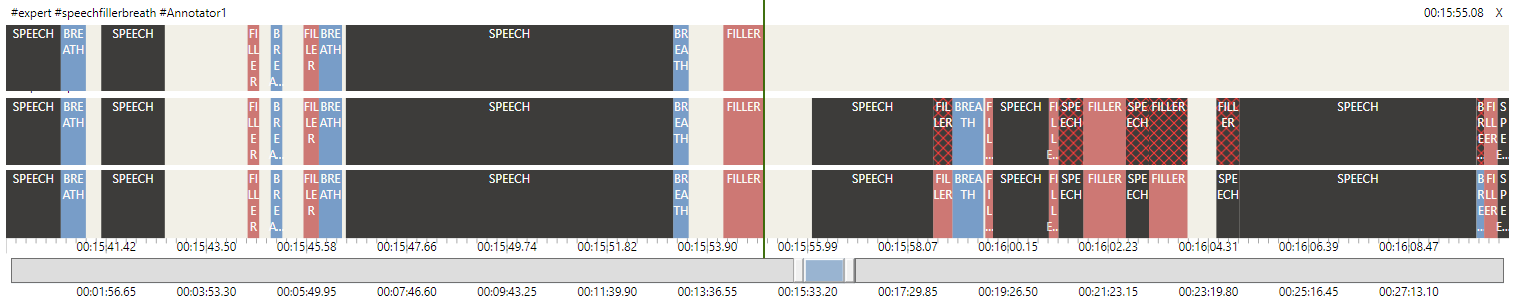}}
\caption{Visualisation of partly finished annotation (upper tier) and the results after the tier is automatically completed (middle tier). Segments with a low confidence are marked with a red pattern. The lower tier shows the final result after manual correction.}
\label{fig:nova:completion}
\end{figure}

To predict annotations, both CML strategies from Section \ref{sec:twofold} are available. In case of \emph{Session Transfer} a dialogue similar to the one in Figure \ref{fig:nova:completion} is shown. However, this time we select a previously trained model and use it to predict the selected sessions. In case of the \emph{Session Completion} step, the annotation is completed by temporarily training a model using only the labels available from current tier. An example before and after the completion is shown in Figure \ref{fig:nova:completion}. The screenshot shows that labels with a low confidence are highlighted with a pattern. This way crucial parts are quickly found and can be revised if necessary.

To assess the prediction accuracy of a model, a dialogue similar to Figure \ref{fig:nova:training} is available. Here, we can pick a trained model and the sessions we want to use for evaluation (only sessions with an according annotation are listed). The model is now applied to predict labels for the selected sessions and the output is compared to the existing annotations. The result is presented in form of a confusion matrix as shown in Figure \ref{fig:nova:confmat}. A confusion matrix provides information on the overall recognition performance, as well as, accuracies for individual classes and which class pairs are often confused.

\begin{figure}[t]
\centering
\fbox{\includegraphics[scale=0.5]{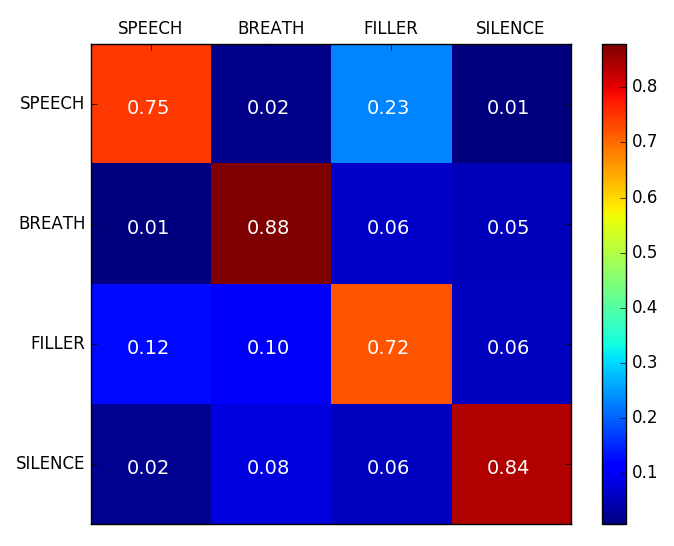}}
\caption{A confusion matrix provides information about the recognition accuracy of individual classes and to what extent they are confused with other classes. For instance, here we see that speech frames are often falsely classified as fillers and vice versa. Hence, an annotator should put attention to these classes while revising the predictions.}
\label{fig:nova:confmat}
\end{figure}


\section{Experiences and Discussion}\label{sec:experiences}

In Section \ref{sec:evaluation} we have presented a technical evaluation of our proposed Cooperative Machine Learning (CML) strategy. Results show that CML bears great potential to significantly reduce human labelling effort. However, it does not necessarily mean that results gained in a simulation can be transferred to human labellers without further ado. Hence, in the following, we want to discuss the experiences of users who have been applying the CML strategies with the \textsc{NOVA} tool introduced in the previous section.

\subsection{What Gain to Expect?}

So, what exact gain can we expect when giving a tool like \textsc{NOVA} into the hands of human labellers? Unfortunately, a general answer to this question does probably not exist. Our experiences show that the amount of time we may save depends on a couple of variables, which may vary from one case from one case to another.

Probably, the largest uncertainty comes from the nature of the annotation problem itself and the ability of the applied machine learning (ML) techniques to cope with it. For instance, let us assume the task of labelling voiced parts in audio. If the recordings have low background noise and speech is really the only prominent signal, a simple feature like loudness may already allow us to train a robust model on few samples, yet generalizing well on unseen data. In this case, the time saving (compared to a completely manual approach) can be tremendous. On the other hand, if the speech files are noisy and contain other audible sounds -- possibly overlapping with speech -- the problem becomes immediately harder. As a consequence not only a more sophisticated feature set (and classification model) is needed, but more manual labelling effort is required to obtain a robust model. As a consequence, less time is saved. And we may even reckon the case where the problems becomes too hard to train a reliable model at all, so that the effort to manually revise the prediction may eat up initial savings. The possibility to exchange features and classifiers in \textsc{NOVA} is therefore an essential precondition to adapt to the problem at hand in the best possible way.

Another point to consider is the quality of the annotation that is desired. Can we live with some false prediction? Or do we aim for a high precision, yet do not mind a high number of false negatives? This, of course, depends very much on the purpose the data is labelled for. As a special flaw social signals often lack a ground truth. And when multiple raters are employed the agreement often turns out to be low. This makes it specially difficult to estimate the quality of a prediction. In the end, it depends a lot on the assessment of the user if he or she is pleased with the automatic completion. Here, \textsc{NOVA}'s feature to immediately visualize the results is an important tool to let raters assess the quality of automatic predictions.

Finally, comparing manual with semi-manual annotations is not as straight forward as it may seem. When observing automatic predictions we observed that on- and offset of the labels were often more precise than that of humans, which are usually rather fuzzy (unless they work at a very fine granular time scale, which is usually too time-consuming). Likewise, we found that short occurrences of a behaviour are easily overlooked by human labellers, especially as their attention drops with time. Hence, since machines show no signs of fatigue their predictions are often more consistent throughout a corpus compared to those of humans. Consequently, applying CML strategies may not just help saving time, but also lead to more accurate and stable annotations.

\subsection{Experienced Annotators}

To learn more about the general applicability of our approach, we asked the experienced annotators we had hired for the manual annotation of the NOXI database (see Section \ref{sec:database}) to redo some of their sessions. The task was again to annotate speech and filler events. This time, however, we explicitly told them to make use of the integrated CML tools. Afterwards we compared the new semi-automated annotation with the previous manual ones and also asked for their subjective impressions.

First of all, they reported that they were positively surprised by the accuracy of the automatically generated labels. Only little manual efforts were required to correct the predictions for the given task. In particular, they found that in some cases detections were even more precise compared to their previous manual annotation. We explain this with the fact that the human brain naturally filters out information it perceives as not relevant in the current context. For instance, we do not consciously hear short breathing sounds during an utterance since we concentrate on the content of the spoken message. And even if we force ourselves to pay special attention to certain events it may be too exhausting to accurately label each and every occurrence. This is a situation in which the semi-automated annotation can really pay off as machines -- in contrast to humans -- do not get tired when they have to repeat the same task over and over again. Hence, behaviour that is easy to detect but occurs too frequently to justify manual efforts should be labelled automatically.

Of course, the precise working method of a machine may not always have the desired effect. The annotators noticed that sometimes incorrect labels were introduced, too. In particular, when filler events were falsely detected at the beginning and ending of an utterance. We explain this with the fact that some fillers are indeed words like ``yeah'' and ``okay'' or at least have a very similar sound (\eg `uh-huh'' and ``hmm''). The classifier learns to label these sounds as fillers if they are surrounded by silence, which is the case at sentence boundaries. Here, often the semantic context \cite{Baur2016} is required to decide whether a word is a filler or part of speech. This is a situation where the automated approach is likely to fail. However, even in that case it can still help to speed up the annotation process since it is usually faster to correct a wrongly assigned label than creating it from scratch (in \textsc{NOVA} hotkeys are available for this purpose).

\subsection{Inexperienced Users}

To see how inexperienced users cope with \textsc{NOVA}, we asked students in an introductory lecture on human-computer interaction to solve an annotation task and fill out a questionnaire afterwards. Firstly, the 14 students were divided into four groups (3,3,4,4) and a quick introduction to \textsc{NOVA} was given. Again we stick to recordings from the \textsc{NOXI} database, but simplified the annotation task to two classes: \textsc{SPEECH} and \textsc{LAUGHTER}. We then asked the students to load one of the sessions and create an empty annotation. After annotating few speech and laughter chunks they could use the session completion tool to finish the remaining part of the session. After observing some of the predicted labels they could then decide to either add more manual labels and repeat the completion step, or revise predicted labels with a low confidence. Finally, we provided a manual annotation of the session and asked them to compare it to their own semi-manual annotation. Each group successfully finished the task in less than an hour.

In the questionnaire we wanted to know what they believe are the strengths and weaknesses of human versus machine coding. We also asked them open questions on how to improve the system. Interestingly, they observed that machine labels were generally more precise, but failed in specific situations \eg when speech and laughter occurred at the same time. Despite the short time they spent with the tool, they already reported a loss of concentration and noted that this does not apply to machines. Regarding the visual guidance during the revision of a prediction, all groups agreed that highlighting labels with a low confidence helped them correct their annotations. However, we were also interested in their opinion on the visualization of this information. Currently, labels below a adjustable confidence threshold are superimposed with an uniform pattern. Such a binary decision has the advantage that the user can quickly detect spots that require actions. On the other hand, it is not evident whether a label is weakly or strongly accepted / rejected. While two of the groups liked the binary highlighting, the other two groups preferred a more detailed visualization, \eg using a colour gradation. One group also mentioned that probabilities for all classes should be available to get a better understanding why a prediction failed. Further investigations are needed to understand if a more finely graduated representation into several confidence classes is preferable.

Generally, the students reported that they had no difficulties using the interface of \textsc{NOVA} and that the integrated CML tools helped them complete the task in less time.

\subsection{Generalisability and Adaptation}

We also investigated how our approach performs with respect to other modalities than audio. To this end, we applied \textsc{OpenFace} \cite{Baltruvsaitis:2016} to extract visual features from the videos in the \textsc{NOXI} database (German sessions). The result is a 196 dimensional feature vector per frame (25 Hz), including facial landmarks, action units and gaze directions. Based on these features facial behaviour can be learned. For the following experiment, we defined the task of smile annotation. Hence, an annotation scheme containing a single label SMILE was applied. We employed an experienced annotator who has been working with \textsc{NOVA} before and introduced him to new CML tools. However, this time we did not set a fixed procedure as we were interested in seeing how he applied the tools to solve the problem in an explorative process. Therefore, we asked him to take notes about his experiences.

As we would expect, the annotator started to apply the session completion step after labelling smiles within the first two minutes of a session. He noted that at first the system was not able to reliably predict the smiles for the remaining session. He therefore, corrected another two minutes of the predicted smiles, removed all the predictions beyond that point and applied the completion step once again. This procedure was repeated until the prediction looked stable so that only few smiles with a low confidence had to be revised. Once the first session was finished, he trained a model and applied it to predict the smiles of the second session (session transfer step) and so on. If the prediction of a new session looked reliable, he completed the session by revising labels with a low confidence. However, for some subjects he noted that the prediction was not stable enough (possibly because no subject with similar facial expressions had been seen by the model yet) and so he decided to apply the session completion step instead. In any case, after completing a session he retrained the model including the new labels. Since the robustness of the models improved with each new session which was added to the training set, the predictions got more accurate towards the end of the corpus. This increasingly speeded up the coding process. In total, it took him less than 6 h to finish the 18 sessions (in comparison, manual filler annotation in Section \ref{sec:results} took more than 14 h).

\subsection{Finding the Sweet Spot}

The previous experiment also shows how it is possible to detect the moment when it is safe to hand the labelling task over to the machine (see the \emph{sweet spot} discussion in Section \ref{sec:implementation}). We noted that with time, a human annotator learns whether it is worth correcting the predicted annotations or instead adding more labels first before letting the machine complete the session. Especially at earlier stages, sometimes a model trained on few subjects may not perform well enough for unseen users. In the latter case, it may be better to continue using the session completion step. Though, there is no automatic way to predict whether session completion or session transfer should be used, the interface of \textsc{NOVA} allows it to quickly explore both options and pick the more promising approach. Either way, with each completed session the training set incrementally grows, improving the robustness and generalisability of the model.

At some point, when enough sessions are available, the user can apply the following strategy to assess the quality of the prediction. Train a model on a subset of the completed sessions and evaluate it on the remaining ones. The obtained confusion matrix (see Section \ref{sec:nova:walkthrough}) provides feedback about the reliability of the labels. For instance, if a class is often confused with another it may be worth to review all predictions of that class, whereas labels predicted with a high confidence may be safely skipped. Generally, visually reviewing predictions in \textsc{NOVA} is key to find an optimal work-flow with respect to a specific task.

\subsection{Summary}

The core idea behind cooperative machine learning (CML) is to create a loop, in which humans start solving a task (here labelling social signals) and over time a machine learns to automatically complete the task. In conventional approaches, this involves at least two parties: an end-user, who has knowledge about the domain, and a machine learning practitioner, who can cope with the learning system. However, to make the process more rapid and focused, \citet{Amershi:et:al:2014} demand that more control should be given to the end-user. To this end, our tool combines a traditional annotation interface with CML functions that can be applied out of the box requiring no knowledge on machine learning. We found it important to give coders the possibility to individually decide when and how to use them in the labelling process. And to assess the reliability of automatic predictions immediate visual feedback is provided, which gives annotators the chance to adapt their strategies at times. By interactively guiding and improving automatic predictions, an efficient integration of human expert knowledge and rapid mechanical computation is achieved. The reported experiments show that even end-users with little or no background in machine learning are able to benefit from the described machine-aided techniques.

We also observed that CML strategies not only have the potential to speed up coding, but can also have a positive influence on the annotator's coding style. Because of the preciseness machine-aided techniques introduce into the coding process the level-of-detail is improved while at the same time human efforts are reduced. Here, strategies to guide the attention of the annotator during inspection of the predicted labels become a crucial matter. As mentioned before \citet{Rosenthal2010} investigated which kind of information should be provided to the user to minimise annotation errors. However, in their studies they concentrate on single images whereas in our case we deal with continuous recordings. To not overload the annotator with too many details we decided to uniformly highlight labels below an adjustable confidence threshold. Our simulations in Section \ref{sec:evaluation} suggest that this approach helps to significantly reduce labelling efforts. However, the exact gain depends highly on the nature and complexity of annotation problem, the applied machine learning techniques, and not least the expertise and subjective attitude of the human coder.


\section{Conclusion}\label{sec:conclusion}

The goal of the presented work is to foster the application of \emph{Cooperative Machine Learning} (CML) strategies to speed up annotation of social signals in large multi-modal databases. Well described corpora that are rich of human behaviour are needed in a number of disciplines, such as Social Signal Processing and Behavioural Psychology. However, populating captured user data with adequate descriptions can be an extremely exhausting and time-consuming task. To this end, we have presented strategies and tools to distribute annotation tasks among multiple human raters (to bundle as much human efforts as possible) and automatically complete unfinished fractions of a database (to reduce human efforts where possible).

In particular, we have proposed a two-fold CML strategy to support the manual coding process (Section \ref{sec:cl}). Applied to a fresh database it first concentrates on completing few individual sessions. A relatively small amount of labels is sufficient to build a session-dependent model, which -- though not strong enough to generalise well across the whole database -- can be used to derive local predictions. Afterwards, a session-independent classification model is created to finish the remaining parts of the database. During both steps, confidence values are created to guide the inspection of the predictions.

To prove the usefulness of the CML approach, we have presented results for a realistic use-case based on a database featuring natural interactions between human dyads. For our experiments in Section \ref{sec:evaluation} we picked the task of detecting fillers in speech. Fillers are an important cue if one aims to study turn taking and interruption strategies. A fast and general audio detection system in combination with a linear classification model has been applied to more than 10 h of natural conversations yielding an average recognition performance of almost 80 \% (four classes: speech, breath, filler and silence). In a simulation we proved that labelling efforts can be significantly reduced using the proposed system. If applied in combination with a revision of instances with a low confidence value, manual inspection was reduced to $\frac{5}{8}$ of the database. In our case, this corresponds to a saving of approximately 2.5 hours (4.1 h instead of 6.6 h).

It was important to us to bring the proposed approach into application. To this end, in Section \ref{sec:nova} we introduced \textsc{NOVA}\footnote{\url{http://github.com/hcmlab/nova}} -- an open-source tool for collaborative and machine-aided labelling. Other than conventional annotation tools \textsc{NOVA} supports a fully collaborative workflow and allows it to distribute annotation tasks among multiple raters. The discussed CML strategies have been integrated and can be directly applied from the interface to speed up manual annotation. The generalisability of the proposed detection system will enable other researchers to adopt the approach for their own databases and annotation tasks in the future.

Experiences with different groups of users show that the CML approach was also positively perceived from an end-user's point of view who were impressed by the system's accuracy. The feedback we obtained made us aware of different styles of annotation adopted by the end-user and the machine. While a machine is able to annotate social signals much faster and more consistently than humans can do, human raters still bring a better understanding for the application in which the models to be trained will eventually be applied. Furthermore, human raters do not just look at the behaviours to be labelled, but also reason about the context in which they occur. Being presented with the results of an automated labelling process might influence human labellers in a positive manner. Nevertheless, one should be aware of the risk that a machine-like style of annotation might not always result in better systems. This is in particular true when social signals are analysed where raters usually disagree on the labels and no objective ground truth can be established. In order to benefit from the complementary skills of machines and human raters, annotation tools like \textsc{NOVA} are needed that aim for a smooth integration of human intelligence and resources.

Overall, our experiments demonstrated the potential of the CML approach in reducing human labour during the annotation process. Future work will focus on the question of how to further leverage the complementary skills of human and machines. The employment of the CML approach requires end-users to incrementally inject information into the training process until a desired system behaviour is achieved. Such a workflow necessitates a tight coordination of machine and human tasks. In particular, it would be desirable to provide end-users with guidelines on when to hand over annotation jobs to the machine. In the paper, we presented statistics to evaluate the injection of new knowledge by the human into the training process.

In our future work, we also plan to extent the current workflow by automatically generating recommendations in which order sessions in a database should be processed. \citet{Poignant2016} suggest the use of hierarchical clustering to select prototypical examples and prioritise them during the coding process. However, it is not straightforward to adapt their techniques to continuous recordings. Alternatively, in our case we can make use of the confidence values generated during label prediction. Using the average value the following strategy is conceivable: every time a session is finished, a model is built to predict remaining sessions and pick the one with the lowest score to complete next. This way we ensure that manual efforts get spent on data that has a high potential to improve the learner in the next iteration.

\section*{Acknowledgement}
This work is Funded by European Union Horizon 2020 research and innovation programme, grant agreement No 645378 (ARIA-Valuspa).

\bibliography{cml}



\medskip

\end{document}